\documentclass[nofootinbib,aps,a4paper,letterpaper,superscriptaddress,
twocolumn,times,eqsecnum]{revtex4}
\usepackage{amsmath,amssymb,amsfonts}
\usepackage{graphicx}
\usepackage{booktabs} 
\usepackage{placeins}
\usepackage{hyperref}
\usepackage{eurosym}
\usepackage{epsfig} 
\usepackage{subfigure}
\usepackage{float}
\usepackage{epstopdf}
\usepackage{multirow}
\usepackage{siunitx} 
\usepackage{adjustbox}
\pdfoutput=1

\usepackage{color}
\usepackage{braket}
\usepackage{dcolumn}
\usepackage{bm,url} 
\newcommand{\Mpl}{M_{\rm Pl}}
\usepackage{amsfonts}
\usepackage[usenames,dvipsnames,svgnames]{xcolor}
\usepackage{hyperref}   
\definecolor{oxfordblue}{rgb}{0.0, 0.13, 0.28}
\definecolor{burgundy}{rgb}{0.5, 0.0, 0.13}
\definecolor{darkolivegreen}{rgb}{0.33, 0.42, 0.18}
\definecolor{darkblue}{rgb}{0,0,0.5}
\definecolor{richcarmine}{rgb}{0.84, 0.0, 0.25}
\definecolor{bluer}{rgb}{0.00,0.50,0.75}{}
\hypersetup{colorlinks=true, citecolor=red, linkcolor=blue,
urlcolor = magenta, filecolor=magenta}

\begin{document}
\title{Unified dark sector and Hubble-tension alleviation in scalar-vector-tensor gravity}

\author{Kimet Jusufi}
\email{kimet.jusufi@unite.edu.mk}
\affiliation{Physics Department, State University of Tetovo,
Ilinden Street nn, 1200, Tetovo, North Macedonia}
\author{Amir A. Khodahami}
\email{a.khodahami@shirazu.ac.ir}
\affiliation{Department of Physics, College of Science, Shiraz
University, Shiraz 71454, Iran}
\author{Ahmad Sheykhi}
\email{asheykhi@shirazu.ac.ir}
\affiliation{Department of Physics, College of Science, Shiraz
University, Shiraz 71454, Iran \\ Biruni Observatory, College of
Science, Shiraz University, Shiraz 71454, Iran}
\author{Jackson Levi Said}
\email{jsaid01@um.edu.mt}
\affiliation{Institute of Space Sciences and Astronomy, University of Malta, 
Malta, MSD 2080 \\ Department of Physics, University of Malta, Malta}
\author{Emmanuel N. Saridakis}
\email{msaridak@noa.gr}
\affiliation{Institute for Astronomy, Astrophysics, Space Applications and 
Remote Sensing, National Observatory of Athens, 15236 Penteli, Greece}
\affiliation{Departamento de Matem\'{a}ticas, Universidad Cat\'{o}lica del 
Norte, Avda. Angamos 0610, Casilla 1280, Antofagasta, Chile}
\affiliation{CAS Key Laboratory for Research in Galaxies and Cosmology, 
School of Astronomy and Space Science,
University of Science and Technology of China, Hefei 230026, China}

\begin{abstract}
We investigate a scalar-vector-tensor theory in which matter is
minimally coupled to a Jordan-frame metric
$\tilde g_{\mu\nu}=(1+\Xi)g_{\mu\nu}$, while a massive vector
sector interacts with the baryonic current. We show that the
conformal scalar coupling modifies the physical expansion rate
measured by matter observers, leading to an enhancement of the
Hubble constant inferred at low redshift. We stress, however, that
the Hubble rate is not a conformal invariant, whereas the acoustic
angular scale $\theta_s$ is, and we derive the exact integral
condition that the scalar field evolution must satisfy. We show
that a single-signed scalar velocity cannot satisfy it, and we
construct instead a two-epoch phenomenological evolution which
matches $\theta_s$ exactly while retaining the late-time enhancement,
at the cost of a small pre-recombination shift of the effective
gravitational coupling. We stress that the scalar is not separately
conserved, and that retaining the scalar--matter energy exchange in its
continuity equation supplies a further restriction: positivity of the
scalar energy density bounds the early component to $z_E\lesssim13$.
Notably, the recombination \emph{temperature} is unmodified, since
particle masses are constant in the Jordan frame, only the expansion
rate at that epoch being altered.
The scalar potential naturally acts as a dynamical dark-energy
sector, while the vector sector provides two distinct contributions.
The temporal component, determined algebraically by the baryon
current, yields an apparent matter-like term in the background
expansion that is not a true fluid but rather a manifestation of
the interaction energy. The propagating spatial modes, on the other
hand, form a vector condensate that behaves as a collisionless
pressureless component and can play the cosmological role of cold
dark matter. Hence, the framework connects
scalar dynamics, effective dark-energy evolution, and the
$H_0$ tension within a single setup.
Finally, local
gravitational constraints require a chameleon-type screening
mechanism, whose efficiency we discuss critically.
\end{abstract}

\maketitle

\section{Introduction}

The $\Lambda$CDM cosmological paradigm has achieved remarkable
success in describing a broad range of observations, from the
anisotropies of the cosmic microwave background (CMB) to the
large-scale distribution of matter and the late-time accelerated
expansion of the Universe \cite{Peebles,Copeland}.
Despite this phenomenological success, several observational and
theoretical tensions have emerged in recent years, suggesting that
the standard cosmological paradigm may be incomplete 
\cite{Weinberg,DiValentino:2020zio,Perivolaropoulos:2021jda}.

The most serious of these discrepancies is the Hubble tension,
namely the persistent disagreement between the value of the
present-day Hubble constant inferred from early-universe probes
and the value measured through late-time observations
\cite{Riess2021,Planck2018}. In particular, the latest SH0ES
results yield
$
H_0=73.0\pm1.0\,{\rm km\,s^{-1}\,Mpc^{-1}},
$
\cite{Riess2021},
whereas the Planck 2018 analysis assuming $\Lambda$CDM gives
$
H_0=67.4\pm0.5\,{\rm km\,s^{-1}\,Mpc^{-1}}
$
\cite{Planck2018}. The statistical significance of this
discrepancy has now exceeded the level at which it can be easily
attributed to statistical fluctuations, motivating the exploration
of new physics beyond the standard cosmological framework
\cite{Abdalla:2022yfr,Verde}.

A large variety of approaches has been proposed in order to
address this tension, including early dark energy scenarios  
\cite{Poulin:2018cxd,Sakstein:2019fmf,Gogoi:2020qif,Niedermann:2020dwg,
Murgia:2020ryi,Chudaykin:2020igl,Seto:2021xua,Freese:2021rjq,Karwal:2021vpk,
Herold:2022iib,Bella:2026zuk,Vagnozzi:2023nrq},
interacting dark sectors 
\cite{DiValentino:2017iww,An:2017crg,Yang:2018qmz,Yang:2018uae,Pan:2019jqh,
Pan:2019gop,Amirhashchi:2020qep,Gao:2021xnk,Guo:2021rrz,Yao:2022kub,DiValentino:2019ffd,Pedrotti:2025ccw}, modified 
recombination 
histories 
\cite{Liu:2019awo,Ye:2020btb,Sekiguchi:2020teg,Lee:2022gzh,
Rashkovetskyi:2021rwg,Lynch:2024hzh,Shepelev:2024guu,Mirpoorian:2024fka,
Jedamzik:2025cax,Pedrotti:2026dwj}, entropic considerations
\cite{Basilakos:2023kvk,Yarahmadi:2024lzd,Adhikary:2025khr,Yarahmadi:2025ujq,
Yarahmadi:2024oqv,Li:2025vqt,Leizerovich:2026pfy},
infrared modifications of gravity 
\cite{Khosravi:2017hfi,Nunes:2018xbm,El-Zant:2018bsc,Cai:2019bdh,Yan:2019gbw,
Escamilla-Rivera:2019ulu,Skara:2019usd,Odintsov:2020qzd,Ballardini:2020iws,
Barker:2020gcp,Braglia:2020auw,Adi:2020qqf,Petronikolou:2021shp,Adil:2021zxp,
Nojiri:2022ski,Banerjee:2022ynv,Schiavone:2022wvq,Ren:2022aeo,Montani:2023xpd,
Boiza:2025xpn,Bouhmadi-Lopez:2026dte}, etc (for a review see 
\cite{CosmoVerseNetwork:2025alb}). Nevertheless,
many proposed models either introduce substantial modifications to
the successful early-universe cosmology or require additional
degrees of freedom and parameter tuning that may reduce their
predictive power. This has motivated increasing interest in
mechanisms capable of altering the effective
cosmological expansion perceived by matter observers while leaving
the physics of recombination and Big-Bang Nucleosynthesis
essentially unchanged. It should be stressed from the outset,
however, that purely late-time modifications of the expansion
history are strongly constrained, since they alter the distance to
last scattering while leaving the sound horizon untouched, and
therefore shift the acoustic angular scale
\cite{Vagnozzi:2023nrq,Pedrotti:2025ccw}. We will make this
statement quantitative within the present framework in
Sec.~\ref{subsec:frame}.

At the same time, recent cosmological analyses have also revived
interest in the possibility of evolving dark energy. In
particular, the latest results from the Dark Energy Spectroscopic
Instrument (DESI) \cite{Adam1,Adam2}, when combined with Type-Ia
supernova and CMB datasets, mildly favor departures from a pure
cosmological constant, with best-fit values corresponding to a
time-dependent dark-energy equation of state. Although the current
statistical significance remains limited, these indications
suggest that the late-time cosmological sector may contain richer
dynamics than those encoded in a strict $\Lambda$CDM description.

Scalar-tensor theories provide a natural framework within which
such modifications of cosmology may arise
\cite{CANTATA:2021asi,Nojiri:2010wj,Clifton:2011jh,Capozziello:2011et}. In 
these theories, additional scalar degrees of
freedom can modify the effective gravitational dynamics and alter
the cosmological expansion history. Importantly, viable scalar
theories can remain compatible with local gravity constraints
through screening mechanisms, such as the chameleon, symmetron,
or Vainshtein mechanisms
\cite{Vainshtein:1972sx,Khoury:2003rn, 
Brax:2021wcv,Joyce:2014kja,Burrage:2017qrf, Heisenberg:2018vsk,Khoury:2013yya}. 
This allows scalar fields to
play a cosmological role at large scales while suppressing
observable deviations from general relativity in high-density
environments.

In the present work we investigate a scalar-vector-tensor
framework in which matter couples conformally to a scalar field
through the Jordan-frame metric
$
\tilde g_{\mu\nu}=(1+\Xi)g_{\mu\nu}$,
with $\Xi$ a dimensionless scalar degree of freedom. Within this
construction, the physical cosmological expansion measured by
matter observers differs from the Einstein-frame expansion rate,
receiving an additional contribution from the scalar dynamics.
This feature provides a mechanism capable of
enhancing the Hubble constant inferred at low redshift. We analyze
carefully what this mechanism can and cannot achieve: since the
comoving distance and the sound horizon are both conformally
invariant, so is their ratio $\theta_s$, and the scalar evolution is
constrained by an exact integral condition which we derive and
impose.

Additionally, the scalar sector simultaneously acts as a dynamical dark-energy
component through its self-interaction potential, allowing for
mild deviations from a pure cosmological constant in qualitative
agreement with recent DESI indications. On the other hand, the
vector sector provides two distinct contributions. The temporal
component, determined algebraically by the baryon current, yields
an apparent matter-like term in the background expansion that is
not a true fluid but rather a manifestation of the interaction
energy. The propagating spatial modes, on the other hand, form a vector condensate that behaves as a collisionless
pressureless component and can play the cosmological role of cold
dark matter. Thus, scalar dynamics drive
the accelerated expansion while the vector sector supplies an
additional matter-like contribution to the expansion history.

The paper is organized as follows. In Sec. \ref{sec:model} we present the
scalar-vector-tensor framework and we derive the corresponding
field equations and discuss the conformal coupling between the
Einstein and Jordan frames. In Sec. \ref{sec:cosmo} we investigate the
effective cosmological expansion and the emergence of the
effective matter sector at the background level, and we discuss the
frame dependence of $H_0$ together with the invariant content of the
tension. In Sec. \ref{latetime} we
introduce the phenomenological scalar ansatz, show that a
single-signed one is excluded by the acoustic scale, and construct a
two-epoch generalization. Then, in Sec. 
\ref{Darkenergysector} we
study the dark-energy sector and we discuss the associated
phenomenological implications. In Sec. \ref{sec:screening} we investigate the
screening mechanism and the consistency with local gravitational
constraints. Finally, Sec. \ref{conclusions} is devoted to the conclusions.
Throughout this work we use units $\hbar=c=1$ and metric signature
$(-,+,+,+)$, and we denote by $\Mpl^2\equiv(8\pi G)^{-1}$ the reduced Planck
mass.

\section{Scalar-vector-tensor framework and conformal coupling}
\label{sec:model}

Scalar-vector-tensor theories of gravity constitute a broad class of
extensions of General Relativity in which the gravitational sector contains,
in addition to the standard massless spin-2 graviton, extra scalar and vector
degrees of freedom (see e.g.~\cite{Moffat:2005si,Benisty:2018qed,
Benisty:2018oyy,Benisty:2021cin}). Such
frameworks can lead to nontrivial cosmological dynamics at late times, while
still recovering standard gravity in appropriate limits.

In this work we consider a scalar-vector-tensor model formulated in the
Einstein frame, described by the action
\begin{equation}
\begin{split}
S = \int d^4x \sqrt{-g}\Big[&
\frac{R}{16\pi G}
- \frac{M^2}{2}\,\frac{(\nabla\Xi)^2}{(1+\Xi)^2}
- V(\Xi)
\Big]\\
&+S_v+\mathcal{I}_m,
\end{split}
\label{eq:action}
\end{equation}
where $\Xi$ is a dimensionless scalar field, $g_{\mu\nu}$ is the
Einstein-frame metric, $V(\Xi)$ is the scalar potential and $M$
is a constant mass scale, which normalizes the scalar kinetic term.
The scale $M$ is required on dimensional grounds: since $\Xi$ is
dimensionless, $(\nabla\Xi)^2$ carries mass dimension two, whereas the
Lagrangian density must carry mass dimension four. As we show in
Sec.~\ref{subsec:limits}, $M$ is not a redundant parameter but fixes the
strength of the scalar-matter coupling relative to gravity, the limit
$M\to\infty$ corresponding to complete decoupling of the scalar sector.

The matter sector is given by
\begin{equation}
\mathcal{I}_m =
S_m[\psi_m,(1+\Xi)g_{\mu\nu}],
\end{equation}
with $\psi_m$ collectively denoting the matter fields.
Moreover, the vector-field contribution is
\begin{equation}
S_v =
\int d^4x \sqrt{-g}\Bigg[
-\frac14 F_{\mu\nu}F^{\mu\nu}
-\frac12 m_v^2 A_\mu A^\mu
\Bigg]
+S_{\rm int},
\end{equation}
where
$
F_{\mu\nu}
=
\nabla_\mu A_\nu-\nabla_\nu A_\mu,
$
while the interaction term between baryonic matter and the vector field reads
\begin{equation}
S_{\rm int}
=
-\int d^4x \sqrt{-g}\,
g_v A_\mu J_b^\mu,
\qquad
J_b^\mu=n_b u^\mu,
\label{eq:Sint}
\end{equation}
with $g_v$ a dimensionless coupling constant, $u^\mu$ the fluid
four-velocity normalized in the Einstein frame ($g_{\mu\nu}u^\mu u^\nu=-1$),
and $n_b$ the Einstein-frame baryon \emph{number} density. We stress that the
source of the vector field is the baryon number current rather than the baryon energy density.
The two descriptions are related by
$n_b=\rho_b/m_b$, with $m_b$ the mean baryon mass; if one prefers to write
$J^\mu_b=\rho_b u^\mu$ then $g_v$ acquires mass dimension $-1$ and must be
interpreted as a charge-to-mass ratio.

The baryon number current satisfies
\begin{equation}
\nabla_\mu J^\mu_b=0,
\qquad
n_b=(1+\Xi)^{3/2}\,\tilde n_b,
\label{eq:current_cons}
\end{equation}
with $\tilde n_b$ the Jordan-frame number density, and taking the divergence 
of the Proca equation and using baryon-number
conservation yields the generalized Proca constraint 
$\nabla_{\mu}\!\left(m_v^2 A^{\mu}\right)=0.$
For a density-dependent vector mass, this does not reduce to the usual
Lorenz condition, but rather gives $ m_v^2 \nabla_{\mu} A^{\mu}
+ A^{\mu}\nabla_{\mu}m_v^2 = 0.$
A current constructed from
the Einstein-frame energy density would not be conserved, precisely because
the conformal coupling exchanges energy between the matter and scalar
sectors. Let us finally note that $S_v$ is defined in the Einstein frame:
although $\sqrt{-g}\,F_{\mu\nu}F^{\mu\nu}$ is conformally invariant in four
dimensions, the Proca mass term is not, and hence the frame assignment of the
vector sector is a physical choice.

The vector mass appearing in $S_v$ is not a constant, but is specified
covariantly in terms of the conserved baryon current. Introducing the scalar
\begin{equation}
n\equiv\sqrt{-J_{b\mu}J^\mu_b},
\qquad
u^\mu=\frac{J^\mu_b}{n},
\label{eq:ncov}
\end{equation}
which coincides with the Einstein-frame baryon number density for a comoving
observer, we take
\begin{equation}
m_v^2(n)=m_\star^2\left(\frac{n}{n_0}\right)^{s},
\label{eq:mvcov}
\end{equation}
with $s$ a dimensionless constant, and $n_0$ the present value of
$n$. Since $n$ is constructed from the current, Eq.~\eqref{eq:mvcov}
defines a scalar function and the action remains completely specified.
Throughout we employ the standard variational formulation of a relativistic
fluid, in which $\sqrt{-g}J^\mu_b$ is metric-independent and
$\partial_\mu(\sqrt{-g}J^\mu_b)=0$ holds, so that all metric
dependence of $n$ is explicit. We anticipate here that two distinct values of
$s$ will play a role below: the interaction component is pressureless for
$s=1$, whereas the vector condensate that can act as cold dark matter
requires the constant-mass case $s=0$. These are properties of different
configurations of the same vector field, and the distinction is developed in
Sec.~\ref{subsec:CMB_vector}; the background numerical analysis uses only the
total vector fraction $\eta$, which is dominated by the condensate.

As we have mentioned, the matter fields are considered minimally coupled to
the conformally rescaled metric
\begin{equation}
\tilde g_{\mu\nu}=(1+\Xi)g_{\mu\nu}\equiv A^2 g_{\mu\nu},
\qquad
A=(1+\Xi)^{1/2},
\label{eq:conformal_trans}
\end{equation}
which defines the Jordan frame. Since matter follows geodesics of
$\tilde g_{\mu\nu}$, observable quantities such as redshift, energy densities,
and the cosmological expansion rate are naturally defined in this frame.
Consequently, the matter energy-momentum tensor is conserved with respect to
$\tilde g_{\mu\nu}$, while the conformal coupling induces an effective exchange
between matter and the scalar sector in the Einstein frame.

The scalar field affects the cosmological dynamics through both its conformal
coupling and its self-interaction potential. In order to simplify the
subsequent analysis, we perform the field redefinition
\begin{equation}
d\varphi=M\,\frac{d\Xi}{1+\Xi}
\qquad\Longrightarrow\qquad
\varphi=M\ln(1+\Xi),
\label{eq:redef}
\end{equation}
where $1+\Xi=e^{\varphi/M}$ which renders the scalar kinetic term canonical. 
The redefined field carries
mass dimension one, $[\varphi]=1$, as a canonically normalized scalar in four
spacetime dimensions must, while the dimensionless combination controlling
every conformal factor is $\varphi/M$. The action then becomes
\begin{equation}
S=
\int d^4x \sqrt{-g}\left[
\frac{R}{16\pi G}
-\frac12(\nabla\varphi)^2
-V(\varphi)
\right]
+S_v+\mathcal{I}_m,
\label{eq:action_canonical}
\end{equation}
corresponding to a scalar-tensor theory with exponential coupling to matter
and a general scalar potential $V(\varphi)$, with
\begin{equation}
\tilde g_{\mu\nu}=e^{\varphi/M}g_{\mu\nu},
\qquad
A(\varphi)=e^{\varphi/2M}.
\label{eq:coupling_function}
\end{equation}
Whenever an explicit potential is required, it must be written with a
dimensionless exponent.

Variation with respect to the metric yields the generalized Einstein equations
\begin{equation}
G_{\mu\nu}
=
8\pi G
\left[
T_{\mu\nu}^{(m)}
+
T_{\mu\nu}^{(\varphi)}
+
T_{\mu\nu}^{(V)}
\right],
\end{equation}
where the scalar-field energy-momentum tensor is
\begin{equation}
T_{\mu\nu}^{(\varphi)}
=
\nabla_\mu\varphi\nabla_\nu\varphi
-
g_{\mu\nu}
\left[
\frac12(\nabla\varphi)^2
+
V(\varphi)
\right],
\end{equation}
while the vector contribution is
\begin{equation}
T_{\mu\nu}^{(V)}
=
F_{\mu\alpha}F_{\nu}^{\ \alpha}
-\frac14 g_{\mu\nu}F^2
+
m_v^2
\left(
A_\mu A_\nu
-
\frac12 g_{\mu\nu}A^2
\right),
\label{eq:TV}
\end{equation}
with
$
F^2=F_{\mu\nu}F^{\mu\nu}
$ and $
A^2=A_\mu A^\mu$. Both expressions carry mass dimension four, as required. We
emphasize that Eq.~\eqref{eq:TV} is the Proca contribution alone. Since
$m_v$ depends through Eq.~\eqref{eq:mvcov} on a matter variable, the variation
of $m_v^2(n)$ with respect to the metric, together with that of $S_{\rm int}$,
generates additional contributions to the vector energy-momentum tensor. These
are retained in Sec.~\ref{subsec:vectorfluid}, where the equation of state of
the vector sector is derived from the complete action; in particular, the
pressure of the vector sector cannot be read off from Eq.~\eqref{eq:TV}
alone.

On the other hand, variation with respect to the vector field gives
\begin{equation}
\nabla_\nu F^{\nu\mu}
-
m_v^2 A^\mu
=
g_v J_b^\mu,
\label{EMV}
\end{equation}
each term carrying mass dimension three, while variation with respect to the
scalar field yields
\begin{equation}
\Box\varphi
-V_{,\varphi}
=
-\frac{d\ln A}{d\varphi}\,T_m
=
-\frac{1}{2M}\,T_m,
\label{eq:scalar_eom}
\end{equation}
where
$
\Box\varphi=\nabla_\mu\nabla^\mu\varphi,
$
and
$
T_m=g^{\mu\nu}T_{\mu\nu}^{(m)}
$
is the trace of the matter energy-momentum tensor. The scalar
field therefore couples directly to the matter trace through the conformal
coupling, with a strength suppressed by the scale $M$. Equivalently, the
Bianchi identity implies
\begin{equation}
\nabla^\mu T^{(m)}_{\mu\nu}
=
\frac{1}{2M}\,T_m\,\nabla_\nu\varphi,
\label{eq:noncons}
\end{equation}
which is the covariant statement that energy is exchanged between the matter
and scalar sectors in the Einstein frame, the total energy-momentum tensor
remaining conserved.

\subsection{Limits and the physical meaning of $M$}
\label{subsec:limits}

Writing the conformal factor in the conventional parametrization
$\tilde g_{\mu\nu}=e^{2\beta\varphi/\Mpl}g_{\mu\nu}$ and comparing with
Eq.~\eqref{eq:coupling_function} identifies the dimensionless scalar-matter
coupling
\begin{equation}
\beta=\frac{\Mpl}{2M}.
\label{eq:beta}
\end{equation}
Hence $M$ is a mass parameter of the theory, equivalent to the coupling
$\beta$ of standard chameleon and symmetron constructions. In the absence of
screening, and for a light scalar, the exchange of $\varphi$ between two test
masses renormalizes the effective Newton constant as
\begin{equation}
G_{\rm eff}=G\left(1+2\beta^2\right)=G\left(1+\frac{\Mpl^2}{2M^2}\right),
\label{eq:Geff}
\end{equation}
while the corresponding Eddington parameter reads
\begin{equation}
\gamma_{\rm PPN}-1=-\frac{4\beta^2}{1+2\beta^2},
\label{eq:gammaPPN}
\end{equation}
which is the standard scalar-tensor result
$\gamma_{\rm PPN}-1=-2\alpha_0^2/(1+\alpha_0^2)$ with $\alpha_0^2=2\beta^2$.
Two choices are natural. The first is
$M=\Mpl$, which introduces no new scale beyond those already present and
gives $\beta=1/2$, that is an order-unity fifth force, $G_{\rm
eff}=\tfrac32G$, in unscreened regions; in this case the screening mechanism
discussed in Sec.~\ref{sec:screening} is essential rather than optional. The
second is to treat $M$ as a free parameter: if the scalar were unscreened on
Solar System scales the Cassini bound
$|\gamma_{\rm PPN}-1|\lesssim2\times10^{-5}$ would
require $\beta\lesssim2.2\times10^{-3}$, i.e. $M\gtrsim2\times10^{2}\Mpl$,
whereas with an efficient thin-shell suppression values $M\sim\Mpl$ would
remain viable. In the numerical analysis below we adopt $M=\Mpl$, namely
$\beta=1/2$; we return in Sec.~\ref{sec:screening} to the question of whether
the required thin-shell suppression is in fact achievable for this value.

Finally, we note that in the limit $\varphi\rightarrow0$ the conformal factor
reduces to unity, namely
$
\tilde g_{\mu\nu}\rightarrow g_{\mu\nu},
$
and the model approaches standard $\Lambda$CDM cosmology when
$V(\varphi)=\text{const.}$ with the scalar sector behaving effectively as a
cosmological constant. Independently, for $M\to\infty$ at fixed $\varphi$ one
has $\beta\to0$, the source term of Eq.~\eqref{eq:scalar_eom} vanishes, the
two frames coincide, and General Relativity with a minimally coupled scalar is
recovered. Neither limit is available in a formulation lacking the scale $M$.

\section{Effective cosmological expansion and matter sector}
\label{sec:cosmo}

In this section we proceed to investigate the cosmological implications
of the scenario at hand at the
background level. In particular, we focus on the way the conformally coupled
scalar sector modifies the effective expansion history, while the
vector sector provides an additional effective matter contribution. As we will
see, the dominant cosmological effect relevant for the present work arises
from the scalar-induced modification of the observed Hubble expansion rate.

We consider a spatially flat Friedmann-Lema\^itre-Robertson-Walker geometry
in the Einstein frame, namely
\begin{equation}
ds^2=-dt^2+a(t)^2d\mathbf{x}^2.
\end{equation}
Under the conformal transformation
\begin{equation}
\tilde g_{\mu\nu}=e^{\varphi/M} g_{\mu\nu},
\end{equation}
the Jordan-frame metric becomes
\begin{equation}
d\tilde s^2
=
e^{\varphi/M}
\left(
-dt^2+a(t)^2d\mathbf{x}^2
\right)
=
-d\tilde t^2+\tilde a(\tilde t)^2d\mathbf{x}^2,
\label{eq:metric-conformal-trans}
\end{equation}
where
\begin{equation}
d\tilde t=e^{\varphi/2M}dt,
\qquad
\tilde a=e^{\varphi/2M}a.
\label{eq:Ein-Jor-relation}
\end{equation}

Therefore, the cosmological expansion rate measured by matter observers is
\begin{equation}
\tilde H\equiv
\frac{1}{\tilde a}
\frac{d\tilde a}{d\tilde t}.
\end{equation}
Using Eq.~\eqref{eq:Ein-Jor-relation}, one obtains
\begin{equation}
\tilde H
=
e^{-\varphi/2M}
\left(
H+\frac{\dot\varphi}{2M}
\right),
\label{eq:Heff_exact}
\end{equation}
where $H=\dot a/a$ is the Einstein-frame Hubble parameter. This relation
shows that the observed cosmological expansion receives contributions both
from the standard background evolution and from the scalar-field dynamics,
the latter being controlled by the dimensionless ratio $\dot\varphi/(MH)$
rather than by $\dot\varphi$ alone.

In order to parametrize the scalar contribution in a convenient way we adopt as
independent variable the \emph{observable} redshift. Since photons and matter
follow geodesics of $\tilde g_{\mu\nu}$, this is defined through the
Jordan-frame scale factor,
\begin{equation}
1+z=\frac{\tilde a_0}{\tilde a}
=e^{(\varphi_0-\varphi)/2M}\,\frac{a_0}{a},
\label{eq:redshift}
\end{equation}
with $\tilde a_0=1$. We stress that $z$ differs from the Einstein-frame ratio
$a_0/a$ precisely by the conformal factor, a distinction that must be retained
since it is of the same order as the effect under study. All density scalings
below, as well as all figures, refer to this physical redshift. We then define
the dimensionless function
\begin{equation}
f(z)\equiv \frac{\dot\varphi(z)}{M H_0},
\qquad
\dot\varphi^2=M^2H_0^2f^2,
\label{eq:f-def}
\end{equation}
where $H_0$ is the present Einstein-frame Hubble parameter. Relation
\eqref{eq:Heff_exact} then becomes
\begin{equation}
\tilde H(z)
=
e^{-\varphi(z)/2M}
\left[
H(z)+\frac12 H_0f(z)
\right].
\label{eq:Heff_final}
\end{equation}

In the regime of interest, namely $|\varphi|\ll M$, the exponential prefactor
remains close to unity, and thus $f(z)$ quantifies the fractional scalar
contribution to the cosmological expansion rate. In particular, a positive $f$
enhances the expansion rate inferred by matter observers relative to the
Einstein-frame one. We adopt throughout the normalization $\varphi_0=0$,
which is not a loss of generality since a constant shift of $\varphi$ is a
pure redefinition of units; with this choice Eq.~\eqref{eq:Heff_final}
evaluated at $z=0$ gives the exact relation
\begin{equation}
\gamma\equiv\frac{H_0}{\tilde H_0}=\left(1+\frac{f_0}{2}\right)^{-1},
\label{eq:gamma_f0}
\end{equation}
so that $\gamma$ is not an independent parameter but is fixed by $f_0$. Only
the field \emph{excursion} $\Delta\varphi$ is physically meaningful, a point
to which we return in Sec.~\ref{subsec:frame}.

\subsection{The vector sector: two distinct contributions}

Before writing the Friedmann equations, it is crucial to clarify the nature
of the vector sector contributions. The massive Proca field contains both
constrained and dynamical degrees of freedom:

\subsubsection*{The temporal (interaction) component}

For the homogeneous cosmological ansatz
\begin{equation}
A_\mu=(A_0(t),0,0,0),
\end{equation}
the field-strength tensor vanishes, $F_{\mu\nu}=0$. The vector field equation
\eqref{EMV} reduces to the constraint
\begin{equation}
m_v^2 A_0 = g_v n_b,
\end{equation}
which determines $A_0$ algebraically in terms of the baryon density:
\begin{equation}
A_0 = \frac{g_v n_b}{m_v^2(n)}.
\end{equation}

Substituting this into the energy density gives
\begin{equation}
\rho_v^{(\rm int)} = \frac12 m_v^2 A_0^2 = \frac12 \frac{g_v^2 n_b^2}{m_v^2(n)}.
\label{eq:rho_interaction}
\end{equation}

We emphasize that this is \emph{not} a true fluid. It is an \emph{apparent}
energy density that arises from the constraint equation for the temporal
component of the vector field. It represents the interaction energy between
the vector field and the baryon current. This component has no independent
dynamics: its evolution is entirely determined by the baryon density. It
does not have its own equation of state or continuity equation in the usual
sense; rather, its behavior is inherited from the baryons through the
constraint.

For the specific choice $m_v^2(n)\propto n$ ($s=1$), this apparent energy
density scales as $\rho_v^{(\rm int)}\propto n_b\propto a^{-3}$, mimicking
a pressureless fluid at the background level. However, at the perturbation
level it is locked to the baryons:
\begin{equation}
\frac{\delta \rho_v^{(\rm int)}}{\rho_v^{(\rm int)}} = \delta_b,
\qquad
\theta_v^{(\rm int)}=\theta_b.
\end{equation}

\subsubsection*{The spatial (condensate) component}

The spatial components $A_i$ are dynamical degrees of freedom
satisfying the Proca equation
\begin{equation}
\ddot A_i + H\dot A_i + m_v^2 A_i - \frac{\nabla^2}{a^2}A_i = 0.
\end{equation}

For a homogeneous configuration and for $m_v\gg H$, these modes undergo
rapid oscillations. After time averaging, they behave as a pressureless
fluid:
\begin{equation}
\rho_v^{(\rm c)} \propto a^{-3},
\qquad
p_v^{(\rm c)} \simeq 0,
\qquad
\delta_v^{(\rm c)} \neq \delta_b.
\label{eq:condensate_properties}
\end{equation}

This is the only true fluid component in the vector sector. It has
independent dynamics, clusters independently of the baryons, and can play
the cosmological role of cold dark matter.

The total energy density from the vector sector is the sum
\begin{equation}
\rho_v = \rho_v^{(\rm int)} + \rho_v^{(\rm c)}.
\label{eq:total_vector_energy}
\end{equation}
Only the second term represents a fluid with independent
perturbations.

\subsection{Friedmann equations}

Under the above considerations, the Einstein-frame Friedmann equation reads
\begin{equation}
H^2
=
\frac{8\pi G}{3}
\left(
\rho_b+\rho_v+\rho_r+\rho_\varphi
\right),
\label{eq:H2_Einstein}
\end{equation}
where $\rho_b$ and $\rho_r$ are respectively the Einstein-frame energy density
of baryons and radiation, and with
\begin{equation}
\rho_\varphi
=
\frac12\dot\varphi^2+V(\varphi)
=
\frac12 M^2H_0^2f^2+V(\varphi).
\end{equation}

For $s=1$, the interaction component becomes
\begin{equation}
\rho_v^{(\rm int)}
=
\eta_{\rm int}\, m_b n_b,
\qquad
\eta_{\rm int}
=
\frac12
\frac{g_v^2 n_{b0}}{m_b\, m_\star^2},
\label{eq:eta_int}
\end{equation}
with $\eta_{\rm int}$ dimensionless. As established in
Sec.~\ref{subsec:CMB_vector}, the total vector fraction entering the
background equations is $\eta=\eta_{\rm int}+\eta_{\rm c}$, and in the regime
of interest it is dominated by the condensate contribution
$\eta_{\rm c}$.

\subsection{Effective fluid description of the interaction component}
\label{subsec:vectorfluid}

As anticipated above, the interaction component is not a true fluid.
However, at the background level, its behavior can be described by an
effective barotropic fluid. Since $F_{\mu\nu}=0$ for the homogeneous ansatz,
$A_0$ carries no kinetic term and is an auxiliary field which may be
eliminated exactly. Using
$A_\mu A^\mu=-A_0^2$ and $A_\mu J^\mu_b=A_0 n$, the relevant Lagrangian density
is
\begin{equation}
\mathcal L_v=\frac12 m_v^2(n)A_0^2-g_v A_0 n ,
\end{equation}
whose stationary point reproduces the constraint $A_0=g_v n/m_v^2(n)$ obtained
above. Substituting back gives the effective Lagrangian
\begin{equation}
\begin{split}
\mathcal L_v^{\rm eff}
&=-\frac12\frac{g_v^2n^2}{m_v^2(n)}
=-\rho_v^{(\rm int)}(n),\\
\rho_v^{(\rm int)}(n)&=\frac{g_v^2 n_0^{\,s}}{2m_\star^2}\,n^{2-s},
\end{split}
\label{eq:Leff}
\end{equation}
so that the energy density coincides with the expression obtained above.

Equation~\eqref{eq:Leff} is precisely the action of a barotropic perfect fluid
constructed from a conserved current, $\mathcal L_v^{\rm eff}=-\rho_v^{(\rm int)}(n)$, and its energy-momentum tensor may now be
obtained by direct variation. Writing $\mathcal J^\mu=\sqrt{-g}J^\mu_b$, which
is metric-independent and conserved, one can show
\begin{equation}
\delta n=\frac{n}{2}\left(u_\mu u_\nu+g_{\mu\nu}\right)\delta g^{\mu\nu}.
\label{eq:deltan}
\end{equation}

Using the above relations one finds 
\begin{equation}
T^{(V)}_{\mu\nu}
=n\frac{d\rho_v^{(\rm int)}}{dn}\,u_\mu u_\nu
+\left(n\frac{d\rho_v^{(\rm int)}}{dn}-\rho_v^{(\rm int)}\right)g_{\mu\nu},
\end{equation}
which is of the perfect-fluid form
\begin{equation}
T^{(V)}_{\mu\nu}=\big(\rho_v^{(\rm int)}+p_v^{(\rm int)}\big)u_\mu u_\nu
+p_v^{(\rm int)}\,g_{\mu\nu},
\end{equation}
 with
\begin{equation}
p_v^{(\rm int)}=n\frac{d\rho_v^{(\rm int)}}{dn}-\rho_v^{(\rm int)}=(1-s)\,\rho_v^{(\rm int)},
\label{eq:wv}
\end{equation}
yielding $w_v^{(\rm int)}=1-s$. The terms noted above, which are absent from Eq.~\eqref{eq:TV}, are precisely
those generated by $\delta n/\delta g^{\mu\nu}$ in Eq.~\eqref{eq:deltan}.
The interaction component therefore behaves as stiff matter for a constant mass
($s=0$, $w_v=1$), which is precisely the case incorrectly obtained from
Eq.~\eqref{eq:TV} alone, and as a pressureless component for
$s=1$, in which case
\begin{equation}
m_v\propto n^{1/2},
\qquad
\rho_v^{(\rm int)}=\eta_{\rm int}\, m_b n\propto a^{-3},
\qquad
p_v^{(\rm int)}=0 .
\end{equation}
The vanishing of $p_v^{(\rm int)}$ is thus a consequence of the complete action
rather than an assumption, and does not follow from the scaling of $\rho_v$ alone.
As a consistency check we note that, since $J^\mu_b$ is exactly conserved in
the Einstein frame, the number density scales as $n\propto a^{-3}$ without any
conformal factor, so that for $s=1$ the interaction component satisfies 
$\dot\rho_v^{(\rm int)}+3H(\rho_v^{(\rm int)}+p_v^{(\rm int)})=0$. It therefore
behaves as an \emph{uncoupled} dust fluid in the Einstein frame, in contrast
with the baryons, whose Einstein-frame density $\rho_b\propto e^{\varphi/2M}a^{-3}$
is not separately conserved because of the conformal coupling.
We finally note that the elimination of $A_0$ is exact for the homogeneous
background, where $F_{\mu\nu}=0$, and remains a good approximation for
inhomogeneities on scales well below the vector Compton wavelength,
$k\ll m_v$.

Finally, we stress the limitations of this identification. Because $\rho_v^{(\rm int)}$ is
tied algebraically to the baryon number density, its perturbations obey
$\delta\rho_v^{(\rm int)}/\rho_v^{(\rm int)}=\delta_b$ for $s=1$: the component is
exactly comoving with, and locked to, the baryons, and does not constitute an
independently clustering component. In particular, it would participate in the
pre-recombination baryon-photon acoustic oscillations rather than supplying
the non-oscillating potential wells required by the observed CMB peak
structure. The identification of $\eta_{\rm int}\tilde\rho_b$ with an effective
dark-matter contribution is therefore to be understood strictly at the level
of the background expansion, and we do not claim equivalence with cosmological
cold dark matter. That role is played by the vector condensate, which we
describe below.

Since matter is conserved in the Jordan frame, the corresponding energy
densities evolve as
\begin{equation}
\tilde\rho_b\propto \tilde a^{-3},
\qquad
\tilde\rho_r\propto \tilde a^{-4}.
\end{equation}
Using the conformal transformation properties of the energy-momentum tensor,
$\rho^{(E)}=A^4\tilde\rho$, one obtains
\begin{equation}
\rho_b=e^{2\varphi/M}\tilde\rho_b,
\qquad
\rho_r=e^{2\varphi/M}\tilde\rho_r,
\end{equation}
so that $\rho_b\propto e^{\varphi/2M}a^{-3}$ while $\rho_r\propto a^{-4}$
exactly, as expected since radiation is traceless and does not source the
scalar. The interaction energy density is not obtained by a conformal rescaling, being
constructed directly from $T^{(V)}_{\mu\nu}$ in the Einstein frame; since it is
proportional to the baryon number density, $n_b=e^{3\varphi/2M}\tilde n_b$, it
carries a different conformal weight,
$\rho_v^{(\rm int)}=\eta_{\rm int}\,e^{3\varphi/2M}\tilde\rho_b$.
Substituting these relations into Eq.~\eqref{eq:H2_Einstein}, and expressing
the result in terms of $\tilde H$, we obtain
\begin{equation}\notag
\tilde{H}^2 =
\frac{8\pi G}{3}
\Big[
\big(e^{2\varphi/M}+\eta\, e^{3\varphi/2M}\big)\tilde{\rho}_b
+ e^{2\varphi/M}\tilde{\rho}_r
+ \rho_{\varphi}
\Big] e^{-\varphi/M}
\end{equation}
\begin{equation}
+\Big[ \tilde{H} e^{\varphi/2M}\frac{\dot{\varphi}}{M}
- \frac{1}{4}\frac{\dot{\varphi}^2}{M^2}\Big]e^{-\varphi/M}.
\label{eq:H2_Jordan_exact}
\end{equation}
Here $\eta$ denotes the \emph{total} vector fraction, defined relative to the
baryonic rest-mass density through
\begin{equation}
\rho_v=\eta\,m_bn_b,
\qquad
\eta=\eta_{\rm int}+\eta_{\rm c},
\qquad
\eta_{\rm c}\equiv\frac{\rho_v^{({\rm c})}}{m_bn_b},
\label{eq:eta_total}
\end{equation}
with $\eta_{\rm int}$ given by Eq.~\eqref{eq:eta_int}, while $\eta_{\rm c}$ is
fixed by the primordial occupation of the spatial Proca modes rather than
derived. As will be shown in Sec.~\ref{subsec:CMB_vector}, the two pieces carry
the same conformal weight,
$\rho_v^{(\rm int)}=\eta_{\rm int}e^{3\varphi/2M}\tilde\rho_b$ and
$\rho_v^{({\rm c})}=\eta_{\rm c}e^{3\varphi/2M}\tilde\rho_b$, and
$\eta_{\rm c}$ is exactly constant, since both $n_b$ and
$\rho_v^{({\rm c})}$ scale as $a^{-3}$ in the Einstein frame with no conformal
factor. The single $\eta$ term in Eq.~\eqref{eq:H2_Jordan_exact} therefore
already accounts for the condensate, which must not be added as a separate
component; in the regime of interest $\eta\simeq\eta_{\rm c}$.

We stress that the term proportional to $\tilde H\dot\varphi/M$ in
Eq.~\eqref{eq:H2_Jordan_exact} is precisely the one carrying the effect under
study: at the present epoch its fractional contribution is
$\gamma f_0\simeq0.12$ for the parameter values used below, whereas the kinetic
term $\dot\varphi^2/4M^2$ contributes at the few per-mille level. It must
therefore be retained even when the conformal factors are set to unity. Only
if the scalar sector is switched off entirely, $f\to0$, does
Eq.~\eqref{eq:H2_Jordan_exact} reduce to the standard form
\begin{equation}
\tilde H^2
=
\frac{8\pi G}{3}
\left[
\tilde\rho_m
+
\tilde\rho_r
+
V(\varphi)
\right],
\qquad
\tilde\rho_m=(1+\eta)\tilde\rho_b,
\label{eq:Friedmann_simple}
\end{equation}
which is the $\Lambda$CDM reference case and not the scenario at hand.

Now,
the Jordan-frame Friedmann equation can be rewritten as
\begin{equation}\notag
\begin{split}
\Big(e^{\varphi/2M}\tilde{H}-\tfrac{\dot{\varphi}}{2M}\Big)^2
={}&\frac{8\pi G}{3}
\Big[
(e^{2\varphi/M}+\eta e^{3\varphi/2M})\tilde{\rho}_b\\
&\qquad
+ e^{2\varphi/M}\tilde{\rho}_r
\Big]
+\frac{8\pi G}{3}\,\rho_{\varphi} .
\end{split}
\end{equation}
Introducing the normalized Jordan-frame expansion rate
$
\tilde E(z)\equiv \frac{\tilde H(z)}{\tilde H_0},
$
together with $\gamma=H_0/\tilde H_0$ of Eq.~\eqref{eq:gamma_f0},
the above equation can equivalently be written as
\begin{equation}
\left[
\tilde{E}(z)e^{\varphi(z)/2M}
-
\frac{\gamma}{2}f(z)
\right]^2
=
\mathcal A(z),
\label{eq:closure}
\end{equation}
where
\begin{eqnarray}\nonumber
\mathcal{A}(z)
&=&
\Big[e^{2\varphi(z)/M}+\eta\, e^{3\varphi(z)/2M}\Big]
\bar\Omega_{b0}(1+z)^3\\
&+&
e^{2\varphi(z)/M}
\bar\Omega_{r0}(1+z)^4
+
\bar\Omega_{\varphi}(z),
\label{eq:Adef}
\end{eqnarray}
with
\begin{equation}
\bar\Omega_{\varphi}(z)
=
\bar\Omega_{\varphi 0}\,
\frac{\rho_{\varphi}(z)}{\rho_{\varphi 0}},
\label{eq:Omegaphi_z}
\end{equation}
and 
\begin{equation}
\rho_{\varphi}(z)=\frac12 M^2H_0^2 f^2(z)+V(\varphi(z)).
\end{equation}
Here all density parameters entering $\mathcal A(z)$ are normalized to the
\emph{present} expansion rate,
\begin{equation}
\bar\Omega_i(z)\equiv\frac{8\pi G\,\rho_i^{(E)}(z)}{3\tilde H_0^2}
=\mathcal A_i(z),
\label{eq:barOmega}
\end{equation}
while the fractional contributions to the expansion rate, normalized to the
running $\tilde H(z)$, are
\begin{equation}
\Omega_i(z)\equiv \frac{8\pi G}{3\tilde H^2(z)}\,e^{-\varphi/M}\rho_i^{(E)}(z)
=\frac{e^{-\varphi(z)/M}\,\mathcal A_i(z)}{\tilde E^2(z)},
\label{eq:Ob}
\end{equation}
where $\rho_i^{(E)}$ denotes the Einstein-frame energy density of the $i$-th
component and $\mathcal A_i$ the corresponding term in
Eq.~\eqref{eq:Adef}, so that the two normalizations are related by
\begin{equation}
\bar\Omega_i(z)=e^{\varphi(z)/M}\,\Omega_i(z)\,\tilde E^2(z),
\label{eq:norm_relation}
\end{equation}
the conformal factor being essential and the two coinciding at $z=0$, where
$\varphi_0=0$ and $\tilde E=1$. We note that $\bar\Omega_{b0}$ is built from
the Jordan-frame density $\tilde\rho_{b0}$, the factor $e^{2\varphi/M}$ in
Eq.~\eqref{eq:Adef} converting it to the Einstein frame through
$\rho_b^{(E)}=e^{2\varphi/M}\tilde\rho_b$. Explicitly, for the total matter
sector and for the scalar,
\begin{eqnarray}
\Omega_m&=&\frac{\left(e^{\varphi/M}+\eta\,e^{\varphi/2M}\right)
\bar\Omega_{b0}(1+z)^3}{\tilde E^2},
\\
\Omega_{\varphi}&=&\frac{e^{-\varphi/M}\bar\Omega_{\varphi}}{\tilde E^2}.
\label{eq:Ophi}
\end{eqnarray}
We stress that the factor $e^{-\varphi/M}$ in Eq.~\eqref{eq:Ob} is not
optional: it is the overall conformal factor multiplying the right-hand side
of Eq.~\eqref{eq:H2_Jordan_exact}, and omitting it would spoil the closure
of the density parameters. Indeed, dividing Eq.~\eqref{eq:closure} by
$\tilde E^2e^{\varphi/M}$ gives the exact sum rule
\begin{equation}
\sum_i\Omega_i(z)=\left[1-\frac{u(z)}
{2\sqrt{\mathcal A(z)}+u(z)}\right]^{2},
\label{eq:sumrule}
\end{equation}
so that the density parameters add up to unity whenever the scalar velocity
vanishes. The origin of the departure becomes transparent once
Eq.~\eqref{eq:sumrule} is written in closed form. One can further show that 
\begin{equation}
\sum_i\Omega_i(z)
=e^{-\varphi(z)/M}\,\frac{H^2(z)}{\tilde H^2(z)}
=\big[1+\epsilon(z)\big]^{-2},
\label{eq:sumrule_closed}
\end{equation}
where
\begin{equation}
\epsilon(z)\equiv\frac{\dot\varphi}{2MH}
=\frac{u(z)}{2\sqrt{\mathcal A(z)}} ,
\label{eq:epsilon_def}
\end{equation}
that is, the sum rule measures the ratio of the Einstein- to the
Jordan-frame expansion rate.  We stress that the Friedmann constraint itself is
unmodified: in the Einstein frame $\sum_i\Omega_i^{(E)}=1$ holds,
and Eq.~\eqref{eq:sumrule_closed} states only that the $\Omega_i$ of
Eq.~\eqref{eq:Ob}, being Einstein-frame densities normalized to the
Jordan-frame rate $\tilde H$, are all rescaled by the common factor
$(1+\epsilon)^{-2}$, hence no individual component is in deficit. Nevertheless, it is convenient to write an effective geometric contribution via $\sum_i\Omega_i+\Omega_u=1$, with
\begin{equation}
\Omega_u
=\frac{2\epsilon+\epsilon^{2}}{(1+\epsilon)^{2}},
\label{eq:Omega_cross}
\end{equation}
with an interesting analogy that it plays the role that spatial curvature plays in the standard case.  

In particular $\sum_i\Omega_i\to1$ both for $z\to-1$ and for $z\to\infty$,
where $f\to0$, while at the present epoch $\varphi_0=0$ and
$\epsilon_0=f_0/2$, so that Eq.~\eqref{eq:sumrule_closed} reduces to
$\sum_i\Omega_i=\gamma^2\simeq0.88$, with $\gamma$ given by
Eq.~\eqref{eq:gamma_f0}. The present-day departure from unity is thus not an
independent number but is fixed by $f_0$, the same quantity that controls the
offset between $H_0$ and $\tilde H_0$. We note finally that the distinction
matters when the model is confronted with data: the present matter fraction is
$\Omega_{m0}=(1+\eta)\bar\Omega_{b0}=0.279$ in the normalization
\eqref{eq:Ob}, but $\Omega_{m0}^{(E)}=\Omega_{m0}/\gamma^2=0.318$ in the
Einstein-frame one. Distance observables depend only on $\tilde H(z)$ and are
unaffected, but growth-rate comparisons are not, and we shall use
Eq.~\eqref{eq:Ob} throughout.
The subscript ``0'' marks the
present value of a quantity, and we note that the two normalizations coincide
at $z=0$. Distinguishing them is essential: mixing the two within
Eq.~\eqref{eq:Adef} would introduce a spurious factor $\tilde E^2(z)$.
Hence, the normalized expansion rate becomes
\begin{equation}
\tilde E(z)
=
e^{-\varphi(z)/2M}\Big[\sqrt{\mathcal A}
+\frac{u(z)}{2}\Big],
\label{eq:Etilde-final}
\end{equation}
where we have defined
\begin{equation}
u(z)\equiv \gamma f(z).
\label{udef}
\end{equation}

In the numerical analysis, we do not need to specify the scalar potential
$V(\varphi)$ explicitly. Instead, the full redshift dependence of the
scalar-field energy density $\rho_\varphi(z)$ is determined by the
phenomenological ansatz for $f(z)$ together with energy conservation.
Defining the kinetic and potential contributions in the normalization
\eqref{eq:barOmega},
\begin{equation}
    \bar\Omega_{\rm kin}(z) \equiv \frac{8\pi G}{3\tilde{H}_0^2}\,
    \frac{1}{2}M^2H_0^2 f^2(z)
    = \frac{1}{6}\left(\frac{M}{\Mpl}\right)^2\gamma^2 f^2(z),
\label{eq:Okin}
\end{equation}
and
\begin{equation}
    \bar\Omega_V(z) \equiv \frac{8\pi G}{3\tilde{H}_0^2}\,V(\varphi(z)),
\end{equation}
we have
\begin{equation}
    \bar\Omega_\varphi(z) = \bar\Omega_{\rm kin}(z) + \bar\Omega_V(z).
\end{equation}
The equation of state of the scalar field can be
expressed solely in terms of $\bar\Omega_\varphi(z)$ and
$\bar\Omega_{\rm kin}(z)$ without reference to the potential:
\begin{equation}
    w_\varphi(z) = \frac{p_\varphi}{\rho_\varphi}
    = \frac{\bar\Omega_{\rm kin}(z) - \bar\Omega_V(z)}
           {\bar\Omega_{\rm kin}(z) + \bar\Omega_V(z)},
\end{equation}
or 
\begin{equation}
    w_\varphi(z) = \frac{2\bar\Omega_{\rm kin}(z) - \bar\Omega_\varphi(z)}
           {\bar\Omega_\varphi(z)},
\end{equation}
this ratio being independent of the choice of normalization.
We stress that the scalar is \emph{not} separately conserved: because of the
conformal coupling it exchanges energy with matter, as already expressed by
the Bianchi identity \eqref{eq:noncons}. Multiplying the Klein--Gordon
equation \eqref{eq:scalar_eom}, which for a homogeneous field reads
$\ddot\varphi+3H\dot\varphi+V_{,\varphi}=T_m/2M$, by $\dot\varphi$ and using
$T_m=-\rho_b$ for pressureless matter gives
\begin{equation}
    \dot{\rho}_\varphi + 3H(\rho_\varphi + p_\varphi)
    = -\,\frac{\rho_b\,\dot\varphi}{2M},
\label{eq:scalar_continuity}
\end{equation}
with the opposite sign appearing in the matter equation, consistently with
$\rho_b\propto e^{\varphi/2M}a^{-3}$. Two features distinguish
Eq.~\eqref{eq:scalar_continuity} from the standard case: the source term on
the right-hand side, and the fact that the friction coefficient is the
Einstein-frame $H$ rather than $\tilde H$. Only the baryons contribute to
$T_m$, radiation being traceless and the vector condensate being an
uncoupled Einstein-frame fluid. Using  Eq.~\eqref{eq:Etilde-final}, and
$dz/dt=-(1+z)e^{\varphi/2M}\tilde H$, we obtain in terms of redshift
\begin{equation}
    \frac{d\bar\Omega_\varphi}{dz}
    = \frac{6\,\bar\Omega_{\rm kin}(z)\sqrt{\mathcal A(z)}
      +\bar\Omega_b^{(E)}(z)\,u(z)/2}
      {(1+z)\left[\sqrt{\mathcal A(z)}+u(z)/2\right]},
\label{eq:Omegaphi_evolution}
\end{equation}
with $\bar\Omega_b^{(E)}=e^{2\varphi/M}\bar\Omega_{b0}(1+z)^3$ and
$\bar\Omega_{\rm kin}$ given by Eq.~\eqref{eq:Okin}. Using these relations
Eq.~\eqref{eq:Omegaphi_evolution} can be equivalently written in the
familiar equation-of-state form
\begin{equation}
    \frac{d\bar\Omega_\varphi}{dz}
    = \frac{1}{1+z}\,
      \frac{3\left[1+w_\varphi\right]\bar\Omega_\varphi
            \sqrt{\mathcal A}+\bar\Omega_b^{(E)}\,u/2}
           {\sqrt{\mathcal A}+u/2},
\label{eq:Omegaphi_evolution_w}
\end{equation}
where we see that the first
correction comes from the frame factor
$\sqrt{\mathcal A}/(\sqrt{\mathcal A}+u/2)$, while the second due to the conformal
source $\bar\Omega_b^{(E)}u/2$. Equation~\eqref{eq:Omegaphi_evolution_w} is an
identity rather than an independent relation, and since $w_\varphi$ is itself
a function of $\bar\Omega_\varphi$ it is implicit; we therefore integrate
Eq.~\eqref{eq:Omegaphi_evolution} in practice. This is integrated
together with Eq.~\eqref{scalarfield} for $\varphi(z)$. 
The scalar potential is then reconstructed from the solution as
\begin{equation}
    V(\varphi(z)) = 3\Mpl^2\tilde{H}_0^2
    \left[\bar\Omega_\varphi(z) - \bar\Omega_{\rm kin}(z)\right],
\label{eq:potential_reconstruction}
\end{equation}
where we used $\Mpl^2=(8\pi G)^{-1}$, so that
$[V]=2+2=4$ as required. Thus, the numerical analysis determines the full
redshift-dependent $\rho_\varphi(z)$ and the corresponding potential
$V(\varphi)$ self-consistently, without requiring an a priori specification
of the potential form. This approach is analogous to the reconstruction of
the inflaton potential from the inflationary observables, where the dynamics
determine the potential rather than the potential being assumed from the
outset.

In the regime $|\varphi|\ll M$ the conformal factors in
Eq.~\eqref{eq:Adef} remain close to unity and
\begin{equation}
\mathcal A(z)\simeq(1+\eta)\bar\Omega_{b0}(1+z)^3+\bar\Omega_{r0}(1+z)^4
+\bar\Omega_{\varphi0},
\label{eq:Aapprox}
\end{equation}
valid when the scalar kinetic term is subdominant and the potential is slowly
varying. We stress that this approximation concerns only the conformal
factors multiplying the density terms; the term $\gamma f(z)/2$ in
Eq.~\eqref{eq:closure}, which carries the entire effect, is always retained.
For the parameter values adopted below the accumulated field excursion is
$|\Delta\varphi|/M\lesssim0.08$, so that Eq.~\eqref{eq:Aapprox} is accurate at
the few-percent level; the results quoted in Sec.~\ref{subsec:two_epoch} use
the exact expression \eqref{eq:Adef}.

\subsection{CMB acoustic peaks and the vector condensate}
\label{subsec:CMB_vector}

The interaction component's pressureless behaviour at the homogeneous
level does not, by itself, imply equivalence with cold dark
matter at the level of cosmological perturbations. For the particular choice
$s=1$, elimination of the homogeneous temporal component gives
\begin{equation}
\rho_v^{(\rm int)}
=\eta_{\rm int} m_b n_b,
\qquad
p_v^{(\rm int)}=0,
\qquad
\rho_v^{(\rm int)}\propto a^{-3}.
\label{eq:vector_background_dust}
\end{equation}
We stress that the combination $g_v^2n_{b0}/2m_bm_\star^2$ obtained
above is precisely $\eta_{\rm int}$, and not the total vector
fraction.
Since this contribution is algebraically tied to the conserved baryon
number density, its perturbations satisfy
\begin{equation}
\frac{\delta\rho_v^{(\rm int)}}
{\rho_v^{(\rm int)}}
=
\frac{\delta n_b}{n_b}
\equiv \delta_b ,
\label{eq:vector_baryon_locking}
\end{equation}
and the corresponding velocity perturbation is likewise locked to the
baryonic one, $\theta_v^{(\rm int)}=\theta_b$.

The interaction component therefore contributes to the inertia of the
photon--baryon fluid without supplying an independently clustering source.
Its presence modifies the baryon-loading parameter according to
\begin{equation}
R\longrightarrow
R_{\rm eff}
=
(1+\eta_{\rm int})R,
\qquad
c_s^2=
\frac{1}{3(1+R_{\rm eff})}.
\label{eq:Reff_shift}
\end{equation}
Consequently, a large baryon-induced contribution during recombination
would modify the sound horizon and the acoustic-peak structure. This
implies that the interaction component must remain subdominant during the
acoustic era. We note in passing that combining
$\eta_{\rm int}=g_v^2n_{b0}/2m_bm_\star^2$ with the Yukawa amplitude
$\alpha_v$ of Eq.~\eqref{eq:alphav} gives
$\eta_{\rm int}/\alpha_v=2\pi G\rho_{b0}/m_\star^2$, which for any
$m_\star\gg H(z_\star)$ and any fifth-force-compatible $\alpha_v$ is
utterly negligible. The requirement is thus automatically satisfied, and it
reinforces the conclusion that the cosmologically relevant vector
contribution must be the condensate.

This conclusion concerns only the baryon-induced temporal configuration
and should not be interpreted as a statement about the complete Proca
sector. A massive vector field contains three propagating degrees of
freedom in addition to the constrained temporal component. In particular,
the same vector field can support an independently populated oscillating
spatial configuration --- the condensate.

To make this distinction explicit, we decompose the vector field as
\begin{equation}
A_0=\bar A_0+\delta A_0,
\quad
A_i=A_i^{({\rm c})}
+\partial_i\chi
+\delta A_i^{T},
\quad
\partial_i\delta A_i^{T}=0 ,
\label{eq:vector_perturbation_decomposition}
\end{equation}
where $\bar A_0$ denotes the baryon-induced homogeneous temporal
configuration, $A_i^{({\rm c})}$ denotes a possible coherent or
statistically isotropic vector condensate, $\chi$ is the longitudinal
perturbation, and $\delta A_i^T$ contains the transverse perturbations.

We emphasize that the interaction component and the condensate are
different manifestations of the same vector field. The interaction
component arises from the constrained temporal mode $A_0$, which is
algebraically determined by the baryon density. The condensate arises
from the propagating spatial modes $A_i$, which have independent dynamics.
These two behaviors can coexist within the same vector field.

The temporal perturbation $\delta A_0$ is not an independent propagating
degree of freedom. Taking the $\mu=0$ component of the Proca equation gives
a constraint. In Newtonian gauge,
\begin{equation}
ds^2=a^2(\tau)
\left[
-(1+2\Psi)d\tau^2
+(1-2\Phi)d{\bf x}^2
\right],
\end{equation}
and, neglecting metric-source terms for the purpose of displaying the
constraint structure, one obtains in Fourier space
\begin{equation}
\left(k^2+a^2m_v^2\right)\delta A_0
=
(1-s)\,g_v\,a^3\,\delta n_b
+k^2\chi' .
\label{eq:A0_constraint}
\end{equation}
Thus $\delta A_0$ remains constrained, whereas $\chi$ and
$\delta A_i^T$ describe propagating vector degrees of freedom. We note that
for $s=1$ the baryonic source cancels, so that in that case the
constrained temporal perturbation is driven solely by the longitudinal mode.

For the interaction component,
\begin{equation}
\rho_v^{(\rm int)}
=
\frac{g_v^2n_b^2}{2m_v^2(n)},
\end{equation}
and, for
\begin{equation}
m_v^2(n)
=
m_\star^2
\left(\frac{n_b}{n_{b0}}\right)^s ,
\end{equation}
one finds
\begin{equation}
\rho_v^{(\rm int)}
\propto n_b^{\,2-s},
\qquad
w_v^{(\rm int)}=1-s.
\label{eq:induced_general_s}
\end{equation}
The corresponding ratio to the baryonic rest-mass density is
\begin{equation}
\eta_{\rm int}(a)
\equiv
\frac{\rho_v^{(\rm int)}}{m_b n_b}
=
\frac{g_v^2 n_b}
{2m_bm_v^2(n)}
\propto
a^{-3(1-s)} .
\label{eq:eta_ind_evolution}
\end{equation}
Hence $s=1$ gives a constant $\eta_{\rm int}$ and a pressureless interaction
component, whereas other values of $s$ lead to a time-dependent fraction.
More generally one may introduce
\begin{equation}
s_{\rm eff}(n_b)
\equiv
\frac{d\ln m_v^2}{d\ln n_b},
\end{equation}
for which
\begin{equation}
w_v^{(\rm int)}
=
1-s_{\rm eff},
\qquad
\frac{d\ln\eta_{\rm int}}{d\ln n_b}
=
1-s_{\rm eff}.
\label{eq:seff_relations}
\end{equation}
A density-dependent mass can therefore make the interaction component
negligible around recombination while allowing it to become more relevant
at late times. This freedom is, however, restricted. For $s_{\rm eff}>1$ one
has $w_v^{(\rm int)}<0$, so that the interaction component behaves as a
dark-energy-like contribution rather than as matter, and it must then be
checked against the scalar dark-energy sector of
Sec.~\ref{Darkenergysector}, which it would otherwise duplicate. For
$s_{\rm eff}>2$ the interaction energy density grows with time, and for
$s_{\rm eff}=2$ it is exactly constant, corresponding to a cosmological
constant fixed by $g_v^2n_{b0}^2/2m_\star^2$. The range is thus
$1<s_{\rm eff}<2$, within which the suppression at recombination is obtained
at the price of a negative late-time equation of state.

\subsubsection*{Constant-mass sector and vector condensate}

For the condensate to behave as cold dark matter, a particularly simple
possibility arises for a vector state with
\begin{equation}
s=0,
\qquad
m_v=m_\star={\rm const}.
\label{eq:s0_constant_mass}
\end{equation}
The homogeneous temporal equation then gives
\begin{equation}
\bar A_0
=
\frac{g_vn_b}{m_\star^2},
\end{equation}
and therefore
\begin{equation}
\rho_v^{(\rm int)}
=
\frac{g_v^2n_b^2}{2m_\star^2}
\propto a^{-6},
\qquad
w_v^{(\rm int)}=1 .
\label{eq:s0_induced}
\end{equation}
Thus the interaction part of the $s=0$ state is a rapidly redshifting
stiff component and cannot constitute the dominant dark matter. We note that,
since a stiff component grows towards the past relative to radiation, its
amplitude is in this case bounded by the expansion rate at nucleosynthesis
rather than by the acoustic physics.

The spatial Proca modes, however, obey an independent dynamical equation.
For a homogeneous spatial configuration and constant mass, the equation
takes schematically the form
\begin{equation}
\ddot A_i^{({\rm c})}
+H\dot A_i^{({\rm c})}
+m_\star^2 A_i^{({\rm c})}
\simeq0 .
\label{eq:vector_condensate_eom}
\end{equation}
When
\begin{equation}
m_\star\gg H ,
\label{eq:rapid_vector_oscillations}
\end{equation}
the microscopic oscillation time $m_\star^{-1}$ is much shorter than the
Hubble time $H^{-1}$. The vector therefore undergoes many coherent
oscillations during one expansion time,
\begin{equation}
A_i^{({\rm c})}
\simeq
{\cal V}_i(t)
\cos(m_\star t+\delta_i),
\end{equation}
with a slowly varying envelope ${\cal V}_i(t)$.

Averaging over time intervals satisfying
\begin{equation}
m_\star^{-1}
\ll\Delta t
\ll H^{-1},
\end{equation}
gives the virial relation
\begin{equation}
\left\langle
\frac{\dot A_i^{({\rm c})}\dot A_i^{({\rm c})}}{a^2}
\right\rangle
\simeq
\left\langle
\frac{m_\star^2
A_i^{({\rm c})}A_i^{({\rm c})}}{a^2}
\right\rangle .
\end{equation}
Consequently,
\begin{equation}
\left\langle p_v^{({\rm c})}\right\rangle
\simeq0,
\qquad
\dot\rho_v^{({\rm c})}
+3H\rho_v^{({\rm c})}
\simeq0,
\end{equation}
and hence
\begin{equation}
\rho_v^{({\rm c})}
\propto a^{-3}.
\label{eq:vector_condensate_dust}
\end{equation}
Equivalently, in the particle description,
\begin{equation}
\rho_v^{({\rm c})}
\simeq
m_\star n_\psi,
\qquad
n_\psi\propto a^{-3}.
\end{equation}
This is the same physical object invoked in the STVG--MOG analysis
\cite{Moffat2026}, where the nonrelativistic excitations of a massive
gravitational vector provide a collisionless, pressureless component that
leaves Thomson scattering, recombination, baryon loading and photon
diffusion unchanged. We emphasize that such a condensate is dynamically
equivalent to cold dark matter on the scales relevant for the acoustic
peaks; it differs from a conventional dark-matter candidate not in its
cosmological behaviour but in belonging to the gravitational sector of the
action, and its abundance is set by initial conditions rather than derived.
Note that vector coherent oscillation dark matter have been investigated also in Ref. \cite{Kitajima:2023fun,Nakayama:2019rhg}.
It is important to emphasize that $s=0$ does not by itself generate a
nonzero condensate. Rather, the constant-mass Proca equations admit such a
solution if the propagating spatial modes possess a nonzero primordial
occupation. The abundance of this component is therefore fixed by the
initial conditions or by an early-Universe production mechanism. If
$A_i^{({\rm c})}$ and $\dot A_i^{({\rm c})}$ both vanish initially, the
condensate remains absent.

A single homogeneous spatial vector would select a preferred direction.
An FLRW background therefore requires either an isotropic ensemble of
vector modes satisfying
\begin{equation}
\left\langle A_i^{({\rm c})}\right\rangle=0,
\qquad
\left\langle
A_i^{({\rm c})}A_j^{({\rm c})}
\right\rangle
=
\frac{\delta_{ij}}{3}
\left\langle
A_k^{({\rm c})}A_k^{({\rm c})}
\right\rangle ,
\label{eq:isotropic_vector_condensate}
\end{equation}
or an isotropic triad realization.

The condition $m_\star\gg H$ is especially important during the acoustic
era. If the condensate is to behave as cold matter already at
recombination, one requires
\begin{equation}
m_\star\gg H(z_\star),
\label{eq:mass_CMB_condition}
\end{equation}
rather than merely $m_\star\gg H_0$. Under this condition the condensate
has, after oscillation averaging,
\begin{equation}
w_v^{({\rm c})}\simeq0,
\qquad
c_{s,v}^2\simeq0,
\qquad
\sigma_v\simeq0,
\qquad
\delta_v^{({\rm c})}\neq\delta_b ,
\label{eq:vector_cold_limit}
\end{equation}
and can therefore act as an independently clustering source for the metric
potentials during the acoustic era.

The scalar perturbations of this cold vector component then approximately
obey
\begin{align}
\delta_v'
&=
-\theta_v+3\Phi',
\\
\theta_v'
+\mathcal H\theta_v
&=
k^2\Psi ,
\label{eq:vector_perturbations_cdm}
\end{align}
which have the same form as the standard cold-matter perturbation
equations.
We note, however, that a nonvanishing $A_i^{({\rm c})}$ renders the
interaction term $g_vA_iJ^i_b=g_vA_i n_bu^i$ first order in the baryon
peculiar velocity, so that the condensate exchanges momentum with the
baryons. In the presence of a nonzero condensate, the baryon velocity
equation can contain an extra drag term,
\begin{equation}
\theta_b' + \mathcal H\theta_b
=
k^2\Psi
+
a\,\Gamma_{vb}\,\frac{\rho_v^{({\rm c})}}{\rho_b}(\theta_v-\theta_b),
\label{eq:baryon_drag}
\end{equation}
where $\Gamma_{vb}$ is a momentum-transfer rate of mass dimension one,
parametrically $\Gamma_{vb}\sim g_v^2 n_b/(m_b m_\star)$, so that the extra
term carries the same dimension as $\mathcal H\theta_b$; the explicit
coefficient requires the full perturbed Proca system and is not needed for
the present discussion. This term
provides an additional constraint on the coupling $g_v$.
The resulting drag is constrained independently of the CMB by
Lyman-$\alpha$ and satellite-galaxy data, and provides a further upper limit
on $g_v$ which must be verified together with the conditions above. We note
also that for the statistically isotropic ensemble
\eqref{eq:isotropic_vector_condensate} the linear-in-$\theta_b$ drag is
suppressed by the vanishing of $\langle A_i^{({\rm c})}\rangle$, so that
Eq.~\eqref{eq:baryon_drag} should be regarded as an upper estimate.
A detailed treatment of this momentum exchange is left for future work.

\subsubsection*{Implications for the acoustic peaks}

Prior to recombination, photons and baryons form a tightly coupled acoustic
fluid with
\begin{equation}
c_s^2
=
\frac{1}{3(1+R)},
\qquad
R=\frac{3\rho_b}{4\rho_\gamma}.
\end{equation}
Only the baryon-locked interaction component modifies this inertia directly,
through
\begin{equation}
R_{\rm eff}
=
\left[1+\eta_{\rm int}(z)\right]R.
\end{equation}
The independently clustering condensate does not participate in the
photon--baryon acoustic motion. Instead, its density perturbation enters the
gravitational potential,
\begin{equation}
k^2\Phi
\simeq
-4\pi G_{\rm eff}a^2
\left[
\rho_b\delta_b
+\rho_v^{({\rm c})}\delta_v^{({\rm c})}
+\rho_v^{(\rm int)}\delta_v^{(\rm int)}+\rho_\gamma\delta_\gamma
\right].
\label{eq:poisson_vector_cmb}
\end{equation}
The two vector contributions therefore play physically distinct roles:
the interaction component modifies baryon loading, whereas the
oscillating spatial condensate can provide the nonoscillating gravitational
source required to sustain the metric potentials while the photon--baryon
plasma oscillates. We add that the peak structure
cannot be reproduced by a rescaling of the gravitational coupling, since a
constant enhancement of $G_{\rm eff}$ multiplies every term in
Eq.~\eqref{eq:poisson_vector_cmb} alike and therefore does not alter the
decay of $\Phi$ at horizon entry. What is required is an additional
\emph{non-oscillating} source. We stress that this statement concerns the
perturbation equations only; the homogeneous rescaling of $G_{\rm eff}$
induced by a nonzero $\varphi_\star$, discussed in
Sec.~\ref{subsec:two_epoch}, acts on the background expansion and hence on
the sound horizon, which is a logically separate effect.

It is consequently useful to write
\begin{equation}
\rho_v
=
\rho_v^{(\rm int)}
+
\rho_v^{({\rm c})}.
\label{eq:vector_split}
\end{equation}
The corresponding present-day density parameters should likewise be kept
separate,
\begin{equation}
\bar\Omega_{v0}
=
\bar\Omega_{{\rm int},0}
+
\bar\Omega_{{\rm c},0}.
\end{equation}
In the constant-mass case their background evolution is different,
\begin{equation}
\rho_v^{(\rm int)}
=
\rho_{{\rm int},0}a^{-6},
\qquad
\rho_v^{({\rm c})}
=
\rho_{{\rm c},0}a^{-3},
\label{eq:vector_two_scalings}
\end{equation}
so that the two contributions cannot in general be described by a single
constant parameter.

We now return to the parametrization \eqref{eq:eta_total} of the total vector
contribution, $\rho_v=\eta(a)\,m_bn_b$ with
$\eta(a)=\eta_{\rm int}(a)+\eta_{\rm c}(a)$. We emphasize that the
reference density there is $m_bn_b$ rather than the Einstein-frame baryon
energy density $\rho_b=e^{\varphi/2M}m_bn_b$. This choice is what makes the
parametrization exact: as established in Sec.~\ref{subsec:vectorfluid}, the
conservation of $J^\mu_b$ implies $n_b\propto a^{-3}$ with no conformal
factor, so that for a condensate in the rapidly oscillating regime, where
$\rho_v^{({\rm c})}\propto a^{-3}$, the ratio
\begin{equation}
\eta_{\rm c}=\text{const}
\label{eq:eta_c_const}
\end{equation}
holds exactly rather than approximately. Had $\eta$ been referred to
$\rho_b$ instead, it would have acquired a spurious drift
$\propto e^{-\varphi/2M}$.

When the interaction contribution is subdominant one therefore has
$\eta\simeq\eta_{\rm c}$, a constant, and the background matter density
retains the form employed above. It should be emphasized,
however, that this equality concerns the homogeneous background only: the
condensate constitutes an independently clustering component and therefore
does \emph{not} satisfy $\delta_v^{({\rm c})}=\delta_b$. Accordingly, the
value of $\eta$ used in the numerical analysis is to
be understood as $\eta\simeq\eta_{\rm c}$, fixed by the initial population of
the condensate, rather than as the combination $g_v^2n_{b0}/2m_bm_\star^2$,
which determines only $\eta_{\rm int}$.

Because the condensate is an uncoupled dust fluid in the Einstein frame,
$\rho_v^{(\rm c)}\propto a^{-3}$, while $\nabla_\mu J^\mu_b=0$ gives
$n_b\propto a^{-3}$ with no conformal factor, the two vector contributions
carry the \emph{same} conformal weight,
\begin{equation}
\rho_v^{(\rm int)}=\eta_{\rm int}e^{3\varphi/2M}\tilde\rho_b,
\qquad
\rho_v^{({\rm c})}=\eta_{\rm c}\,e^{3\varphi/2M}\tilde\rho_b ,
\end{equation}
so that the Friedmann equation retains its form with the single replacement
$\eta\to\eta_{\rm int}+\eta_{\rm c}$. The value of $\eta$ employed in the
numerical analysis is accordingly to be read as $\eta\simeq\eta_{\rm c}$.

A complete demonstration requires deriving the full perturbation system
from the scalar--vector--tensor action and implementing it in a Boltzmann
solver. In particular, one must determine the effective sound speed and
anisotropic stress of the condensate, its coupling to baryonic
perturbations, the stability of the longitudinal sector, and whether the
conditions
\begin{equation}
m_\star\gg H(z_\star),
\qquad
c_{s,v}^2\ll1,
\qquad
\sigma_v\simeq0
\end{equation}
can be maintained throughout the acoustic era. We therefore do not claim
at this stage that the model reproduces the observed CMB power spectrum.
Rather, the analysis identifies the dynamical conditions under which the
vector sector can provide the independently clustering component required
by the acoustic phenomenology.

\subsection{Frame dependence of $H_0$ and the invariant content of the
tension}
\label{subsec:frame}

Equation~\eqref{eq:Heff_final} makes explicit that the Hubble rate is not a
conformal invariant. Under $\tilde g_{\mu\nu}=A^2g_{\mu\nu}$ the expansion
scalar of the matter congruence transforms as
\begin{equation}
\tilde\theta=A^{-1}\left(\theta+3u^\mu\partial_\mu\ln A\right),
\label{eq:theta_transf}
\end{equation}
so that ``the Hubble constant'' is a property of a metric \emph{together
with} a matter congruence, rather than of the spacetime alone. It is
therefore natural to ask whether the discrepancy between early- and
late-time determinations of $H_0$ might be attributed, at least in part, to
the frame in which each is quoted.

The answer requires identifying what is actually measured. Since the
conformal factor cancels between $d\tilde t$ and $\tilde a$ in
Eq.~\eqref{eq:Ein-Jor-relation}, the comoving coordinate separation along a
null geodesic,
\begin{equation}
\Delta x=\int\frac{d\tilde t}{\tilde a}=\int\frac{dt}{a}
=\int\frac{dz}{\tilde H(z)},
\label{eq:comoving_invariant}
\end{equation}
is frame-independent, the last equality following from
Eq.~\eqref{eq:redshift} and holding in either frame. We stress
that it is the Jordan-frame rate $\tilde H$ that appears, for the same
reason discussed below Eq.~\eqref{eq:phi_reconstruction}: $z$ is the
observable redshift, defined through $\tilde a$. The same cancellation
applies to the sound horizon $r_s=\int c_s\,dz/\tilde H$. Consequently the
acoustic angular scale
\begin{equation}
\theta_s=\frac{r_s(z_\star)}{D_M(z_\star)},
\qquad
D_M(z_\star)=\int_0^{z_\star}\frac{dz}{\tilde H(z)},
\label{eq:theta_s}
\end{equation}
which is the quantity determined by the CMB peak spacing and measured to
$0.03\%$ accuracy \cite{Planck2018}, is a conformal invariant. So is the
intercept of the local distance--redshift relation, since laboratory rulers
are atomic lengths and transform together with the metric.

Throughout we distinguish carefully between two closely related scales. The
CMB acoustic angle involves the sound horizon at photon decoupling,
$r_s\equiv r_s(z_\star)$ with $z_\star\simeq1090$, whereas baryon acoustic
oscillations measure the sound horizon at the end of the drag epoch,
\begin{equation}
r_d\equiv r_s(z_d)=\int_{z_d}^{\infty}\frac{c_s\,dz}{\tilde H(z)},
\qquad z_d\simeq1060 ,
\label{eq:rd}
\end{equation}
which is the larger of the two because the sound wave continues to propagate
between $z_\star$ and $z_d$. For the reference $\Lambda$CDM cosmology used
below we obtain $r_d/r_s=1.018$. Since the modification introduced in this
work rescales $\tilde H$ by an almost constant factor throughout the
pre-recombination era, the two scales are shifted by essentially the same
\emph{fractional} amount, as may be verified from
Table~\ref{tab:two_epoch}; the distinction nevertheless matters when
comparing with BAO data, which we do in terms of $r_d$.

The tension is therefore properly formulated between invariants and cannot
be removed by a redefinition of the frame in which $H_0$ is quoted. This has
a sharp consequence for the present framework. Writing
Eq.~\eqref{eq:theta_s} as
\begin{equation}
\int_0^{z_\star}\frac{dz}{\tilde H(z)}=\frac{r_s(z_\star)}{\theta_s},
\label{eq:integral_constraint}
\end{equation}
with $\theta_s$ measured and $r_s$ determined once the physical densities
$\omega_b$ and $\omega_m$ are fixed by the peak morphology, any admissible
$f(z)$ must satisfy an exact integral condition. Since $f$ enters
$\tilde H$ through Eq.~\eqref{eq:Etilde-final} with a positive sign, a
strictly positive $f(z)$ reduces $D_M$ and therefore \emph{necessarily}
violates Eq.~\eqref{eq:integral_constraint} unless $r_s$ is modified as
well. This is the framework-specific version of the general statement that
purely late-time modifications cannot resolve the $H_0$ tension
\cite{Vagnozzi:2023nrq,Pedrotti:2025ccw}. Physical traction thus requires
either a sign change in $f(z)$, or a shift of $r_s$ itself; both are
realized by the two-epoch ansatz constructed in
Sec.~\ref{subsec:two_epoch}.

\section{Scalar dynamics and the Hubble tension}
\label{latetime}

In this section we specify the evolution of the scalar field in a
phenomenological way. The aim is not to reconstruct the full scalar potential
from first principles, but rather to identify a simple and controlled scalar
evolution that leaves Big-Bang Nucleosynthesis and recombination unaffected,
while enhancing the observed Hubble rate at recent times and respecting the
integral condition \eqref{eq:integral_constraint}.

\subsection{Ansatz for the scalar evolution}

The scalar-field dynamics enter the cosmological expansion through the
dimensionless function
$
f(z)$
introduced in Eq.~\eqref{eq:f-def}. Hence, from
the Jordan-frame expansion rate \eqref{eq:Etilde-final}, we deduce that
specifying $f(z)$, or equivalently $u(z)$, determines the
departure from the standard background evolution.

The scalar evolution should satisfy two basic requirements. First, it should
not introduce sizeable deviations at very early times, in order not to spoil
the successful predictions of Big-Bang Nucleosynthesis. Second, it should
become relevant at low redshifts, where it can modify the inferred
value of the present Hubble rate.

During matter domination one has
\begin{equation}
H(z)\propto (1+z)^{3/2}.
\end{equation}
Therefore, in order for the scalar contribution in
Eq.~\eqref{eq:Heff_final} to induce an approximately constant fractional
correction during this epoch, the function $f(z)$ should exhibit the same
scaling behaviour. In this case the quantity
$
H_0f(z)/2H(z)
$
remains nearly constant at high redshift, avoiding large modifications of the
early expansion history.

Motivated by these considerations, we adopt the phenomenological building
block
\begin{equation}
b(z;\alpha,z_i)
=
\alpha\,x_i^{3/2}\,e^{-x_i^{3}},
\qquad
x_i\equiv\frac{1+z}{1+z_i},
\label{eq:bump}
\end{equation}
which peaks at $x_i=2^{-1/3}$, that is at $1+z=0.794\,(1+z_i)$, with maximum
value $0.429\,\alpha$, and is cut off super-exponentially above $z_i$. The
single-component ansatz considered first is
\begin{equation}
f(z)=b(z;\alpha,z_t),
\label{eq:f_ansatz}
\end{equation}
where $\alpha$ is a positive dimensionless parameter and $z_t$ denotes the
transition redshift to the dark-energy dominated era, typically
$z_t\sim0.5$-$0.7$. 
This ansatz is not intended to arise from a unique fundamental potential.
Rather, it provides a minimal parametrization of the desired physical
behaviour. At intermediate redshifts it follows the matter-era scaling
$f(z)\propto(1+z)^{3/2}$, while at very large redshifts the exponential factor
prevents any unbounded growth. At low redshifts, on the other hand, the
scalar contribution becomes non-negligible and can modify the
expansion history. Finally, the present value of the function is
\begin{equation}
f_0
=
\alpha
(1+z_t)^{-3/2}
\exp\!\left[-(1+z_t)^{-3}\right],
\end{equation}
and therefore
$
u_0=
\gamma f_0$, with $\gamma$ given by Eq.~\eqref{eq:gamma_f0}.

The scalar-field evolution can be reconstructed from
Eq.~\eqref{eq:f-def}, together with the Jordan-frame relation
$
d z/d\tilde t=-(1+z)\tilde H(z),
$
equivalently $dz/dt=-(1+z)e^{\varphi/2M}\tilde H(z)$,
which gives
$
d\varphi/dz
=
-M H_0f(z)/[(1+z)e^{\varphi/2M}\tilde H(z)].
$
Hence,
\begin{equation}
\frac{\varphi(z)}{M}
=
-
\int_0^z
\frac{H_0f(z')}{(1+z')\,e^{\varphi(z')/2M}\,\tilde H(z')}
dz',
\label{eq:phi_reconstruction}
\end{equation}
where we recall the normalization $\varphi_0=0$, and where we note that it is
the dimensionless ratio $\varphi/M$ that is directly
determined by $f(z)$. We emphasize that the expansion rate entering here is the
Jordan-frame one, since $z$ is the observable redshift of
Eq.~\eqref{eq:redshift}; employing the Einstein-frame relation
$dz/dt=-(1+z)H$ instead would be inconsistent and would misestimate
$d\varphi/dz$ by a relative amount $\sim u/2$.
Now, using Eq.~\eqref{eq:Heff_final}, the Einstein-frame Hubble parameter can
be written as
\begin{equation}
H(z)
=
e^{\varphi(z)/2M}\tilde H(z)
-
\frac12 H_0 f(z),
\end{equation}
and in terms of $\tilde E(z)$ and $u(z)$ this becomes
\begin{equation}
H(z)
=
\tilde H_0
\left[
e^{\varphi(z)/2M}\tilde E(z)
-
\frac12 u(z)
\right].
\end{equation}
Thus, we obtain
\begin{equation}
\frac{1}{M}\frac{d\varphi}{dz}
=
-
\frac{u(z)}
{(1+z)\,e^{\varphi(z)/2M}\,\tilde E(z)},
\label{eq:phi_tildeE}
\end{equation}
and finally, substituting Eq.~\eqref{eq:Etilde-final},
\begin{equation}
\frac{1}{M}\frac{d\varphi}{dz}
=
-
\frac{u(z)}
{(1+z)\left[\sqrt{\mathcal A(z)}+\frac{u(z)}{2}\right]}.
\label{scalarfield}
\end{equation}
By construction, for moderate values of $\alpha$ the
scalar field remains perturbatively small over the redshift range relevant
for late-time cosmology. Thus, the ansatz above provides a controlled
deformation of $\Lambda$CDM.

\subsection{Modified expansion history and the acoustic scale}

The impact of the scalar field on the expansion rate follows directly from
Eq.~\eqref{eq:Heff_final}. In the small-field regime, $|\varphi|\ll M$, one
obtains
\begin{equation}
\tilde H(z)
\simeq
H(z)+\frac12H_0f(z),
\label{eq:H_simple}
\end{equation}
or in normalized form $\tilde E(z)\simeq E(z)+u(z)/2$.
Thus, the scalar dynamics gives an additive contribution to the observed
Hubble expansion rate.

The form of $f(z)$ ensures that the correction is strongly suppressed at very
early times. Indeed, for $z_t=0.5$ the exponent in Eq.~\eqref{eq:bump} at
recombination is $-(1+z_\star)^3/(1+z_t)^3\simeq-3.8\times10^{8}$, so that
$f(z_\star)$ vanishes to any conceivable accuracy and the thermal history is
untouched. At late times, however, the decrease of $H(z)$ makes the scalar
contribution relatively more important, leading to a deviation from the
$\Lambda$CDM expansion history.

It is precisely this feature, however, that brings the single-component
ansatz into conflict with Eq.~\eqref{eq:integral_constraint}. Because
$f(z_\star)$ vanishes, the sound horizon $r_s$ is unchanged,
while the low-redshift enhancement of $\tilde H$ reduces $D_M(z_\star)$.
Evaluating Eq.~\eqref{eq:theta_s} numerically for
$\alpha=0.4$, $z_t=0.5$ and the density parameters listed in
Table~\ref{tab:two_epoch}, we obtain $\theta_s$ larger than its
$\Lambda$CDM value by $2.5\%$, roughly eighty times the observational
uncertainty. A single-signed $f(z)$ therefore cannot,
by itself, provide a viable resolution, in agreement with the general
arguments of Refs.~\cite{Vagnozzi:2023nrq,Pedrotti:2025ccw}.

\subsection{Two-epoch ansatz and the acoustic scale}
\label{subsec:two_epoch}

It is instructive to see why the sign of $f$ matters. With $\varphi_0=0$, a
positive $f$ implies $\dot\varphi>0$, so that
$\varphi_\star\equiv\varphi(z_\star)<0$. Evaluating
Eq.~\eqref{eq:H2_Jordan_exact} on a plateau where $\dot\varphi\simeq0$ gives
\begin{equation}
\tilde H^2\simeq\frac{8\pi G}{3}
\left[e^{\varphi_\star/M}(\tilde\rho_b+\tilde\rho_r)
+\eta\,e^{\varphi_\star/2M}\tilde\rho_b\right],
\label{eq:Geff_early}
\end{equation}
that is, an effective gravitational coupling
$G_{\rm eff}=Ge^{\varphi_\star/M}$ during the acoustic era, whence
$r_s\propto e^{-\varphi_\star/2M}$. A negative $\varphi_\star$ therefore
\emph{increases} $r_s$, while the late-time enhancement decreases $D_M$;
the two effects add rather than compensate in $\theta_s$.

We are thus led to a two-epoch generalization, retaining the same functional
family \eqref{eq:bump},
\begin{equation}
f(z)=b(z;\alpha_L,z_L)-b(z;\alpha_E,z_E),
\label{eq:f_two_epoch}
\end{equation}
with $\alpha_L,\alpha_E>0$ and $z_E\gg z_L$. The late component, with
$z_L\simeq0.5$, reproduces the behaviour discussed above. The early
component, with $z_E\sim5$--$12$, drives $\varphi$ downwards towards the past
and leaves $\varphi_\star>0$ frozen in before recombination, thereby reducing
$r_s$ through Eq.~\eqref{eq:Geff_early}. The scalar excursion is
non-monotonic, and the two components are cleanly separated in redshift: the
late bump is below $10^{-100}$ by $z\sim10$, while the early one is
subdominant below $z\simeq0.9$, which is where $f$ changes sign. Both vanish at Big-Bang Nucleosynthesis.

We stress that in this framework the microphysics of recombination is
unmodified. Since matter is minimally coupled to $\tilde g_{\mu\nu}$, particle
masses and atomic binding energies are constant in the Jordan frame, and
$T_\gamma\propto\tilde a^{-1}$; recombination therefore occurs at the same
temperature as in $\Lambda$CDM, and no modification of the atomic physics is
invoked. The ionization history is nevertheless not strictly identical, since
the Peebles equation balances the recombination rate against $H$: a
per-cent-level change of $\tilde H(z_\star)$ displaces $z_\star$ and the width
of the visibility function at the sub-per-mille level. We neglect this
throughout, holding $z_\star=1090$ fixed, and note that it is of the same
order as the modelling systematics quantified in
Table~\ref{tab:two_epoch}. Only the expansion rate at that epoch is altered
at the level relevant here. The mechanism is thus of the
varying-$G_{\rm eff}$ type, closely related to the modified-recombination
proposals of Refs.~\cite{Sekiguchi:2020teg,Mirpoorian:2024fka,
Jedamzik:2025cax}, but obtained without any change to the recombination
calculation itself.

Integrating Eq.~\eqref{scalarfield} together with
Eq.~\eqref{eq:Omegaphi_evolution}, retaining all conformal factors in
Eq.~\eqref{eq:Adef}, and fixing $\eta$ by the closure condition
\eqref{eq:closure} at $z=0$, we find that $\theta_s$ can be matched exactly
while retaining $\tilde H_0=73\,\mathrm{km\,s^{-1}\,Mpc^{-1}}$. The results
are collected in Table~\ref{tab:two_epoch}.

\begin{table*}[htb]
\centering
\setlength{\tabcolsep}{4pt}
\begin{tabular}{l
  D{.}{.}{1.4} D{.}{.}{1.4} D{.}{.}{2.4} D{.}{.}{1.4} D{.}{.}{1.4}
  D{.}{.}{3.2} D{.}{.}{3.2} D{.}{.}{5.1} D{.}{.}{1.6}}
\hline\hline
& \multicolumn{1}{c}{$\alpha_E$}
& \multicolumn{1}{c}{$\eta$}
& \multicolumn{1}{c}{$\varphi_\star/M$}
& \multicolumn{1}{c}{$G_{\rm eff}/G$}
& \multicolumn{1}{c}{$\bar\Omega_\varphi(z_\star)$}
& \multicolumn{1}{c}{$r_s$}
& \multicolumn{1}{c}{$r_d$}
& \multicolumn{1}{c}{$D_M(z_\star)$}
& \multicolumn{1}{c}{$100\,\theta_s$} \\
\hline
$\Lambda$CDM, this calculation
 & \multicolumn{1}{c}{--} & \multicolumn{1}{c}{--} & \multicolumn{1}{c}{--}
 & 1.0000 & \multicolumn{1}{c}{--}
 & 144.66 & 147.32 & 13887.7 & 1.041626 \\
$\Lambda$CDM, Planck parameters, this calc.
 & \multicolumn{1}{c}{--} & \multicolumn{1}{c}{--} & \multicolumn{1}{c}{--}
 & 1.0000 & \multicolumn{1}{c}{--}
 & 144.28 & 146.94 & 13869.6 & 1.040265 \\
$\Lambda$CDM, Planck 2018 \cite{Planck2018}
 & \multicolumn{1}{c}{--} & \multicolumn{1}{c}{--} & \multicolumn{1}{c}{--}
 & \multicolumn{1}{c}{--} & \multicolumn{1}{c}{--}
 & 144.43 & 147.09 & 13869.6 & \multicolumn{1}{c}{1.04110(31)} \\
\hline
SVT, single component, Eq.~\eqref{eq:f_ansatz}
 & 0.0000 & 5.0926 & -0.0802 & 0.9229 & 0.6165
 & 150.96 & 153.74 & 14143.6 & 1.067324 \\
SVT, two epoch, $z_E=12$
 & 1.1916 & 5.5805 & 0.0188 & 1.0190 & 0.1120
 & 142.41 & 145.02 & 13672.0 & 1.041626 \\
SVT, two epoch, $z_E=10$
 & 1.0640 & 5.6536 & 0.0241 & 1.0244 & 0.2844
 & 141.72 & 144.32 & 13605.8 & 1.041626 \\
SVT, two epoch, $z_E=5$
 & 0.8909 & 6.2801 & 0.0669 & 1.0692 & 0.5541
 & 136.23 & 138.70 & 13078.5 & 1.041626 \\
\hline\hline
\end{tabular}
\caption{Sound horizon, comoving distance to last scattering and acoustic
angular scale for the two-epoch ansatz \eqref{eq:f_two_epoch}, with
$\alpha_L=0.4$, $z_L=0.5$,
$\tilde H_0=73\,\mathrm{km\,s^{-1}\,Mpc^{-1}}$, $\omega_b=0.02237$ and
$\bar\Omega_{\varphi0}=0.60$; the remaining background parameters are
$\Omega_{b0}=0.0420$ and $\Omega_{r0}=7.8\times10^{-5}$, the latter
corresponding to $\omega_\gamma=2.4728\times10^{-5}$ and $N_{\rm eff}=3.00$,
and the normalization scale is $M=\Mpl$. Here
$r_s=r_s(z_\star)$ and $r_d=r_s(z_d)$ are defined in Eq.~\eqref{eq:rd} with
$z_\star=1090$ and $z_d=1060$, and $\theta_s=r_s/D_M(z_\star)$; all
distances are in Mpc.
The first row is the $\Lambda$CDM reference cosmology used throughout
($h=0.675$, $\omega_m=0.1424$, no scalar or vector sector, $f\equiv0$). The
second and third rows quantify the accuracy of our approximations: they
compare the integrals used in this work, evaluated at the Planck 2018
best-fit parameters ($h=0.6736$, $\omega_m=0.1430$, $z_\star=1089.92$,
$z_d=1059.94$, $N_{\rm eff}=3.046$), with the values obtained by
Ref.~\cite{Planck2018} from a full Boltzmann treatment of the same cosmology.
The agreement is $0.10\%$ in $r_s$ and $r_d$, better than $0.01\%$ in $D_M$
and $0.08\%$ in $100\,\theta_s$, the residual being attributable to our use of
a fixed $z_\star$ and of the tight-coupling sound speed rather than a solved
recombination history. The remaining rows are the
scalar-vector-tensor model of this paper, with the scalar sector evolved
according to Eq.~\eqref{eq:Omegaphi_evolution}. For each $z_E$ the amplitude
$\alpha_E$ is chosen so that $\theta_s$ coincides with the value in the first
row rather than with the measured one, so that these modelling systematics
cancel between model and reference; $\eta$ then follows from the closure
condition \eqref{eq:closure} and is not adjusted. The column
$\bar\Omega_\varphi(z_\star)$ records the scalar energy density frozen in
before recombination and is subject to the positivity requirement discussed
in Sec.~\ref{subsec:two_epoch}, which excludes $z_E\gtrsim13$. We stress that
the agreement of the last three rows with the measured acoustic scale is
obtained by construction and is not a prediction; what is nontrivial is that
a solution exists at all at fixed $\tilde H_0$, and the value of
$\varphi_\star$ that it requires.}
\label{tab:two_epoch}
\end{table*}

The required field excursion is small: for $z_E=10$ one finds
$\varphi_\star/M=0.024$, corresponding to a $2.4\%$ shift of
$G_{\rm eff}$ at recombination, comfortably within Big-Bang Nucleosynthesis
bounds and consistent with the regime $|\varphi|\ll M$ assumed throughout.

The source term in Eq.~\eqref{eq:Omegaphi_evolution} is not a small
correction here, and it supplies a further constraint that would be missed
if the scalar were treated as separately conserved. Where $f<0$, that is
throughout the early component, the term $\bar\Omega_b^{(E)}u/2$ is negative
and grows as $(1+z)^3$, so that the scalar energy density decreases towards
the past. Requiring $\bar\Omega_\varphi(z)>0$ at all redshifts therefore
bounds the amplitude of the early component. Solving for $\alpha_E$ at fixed
$\theta_s$ and tracking $\bar\Omega_\varphi(z_\star)$, we find that it
vanishes at $z_E\simeq13$ and is negative above, so that the admissible
window is
\begin{equation}
z_E\lesssim13 ,
\label{eq:zE_bound}
\end{equation}
with $\bar\Omega_\varphi(z_\star)$ falling from $0.554$ at $z_E=5$ to
$0.112$ at $z_E=12$. This is a restriction of the parameter space,
absent at the level of the background distances alone, and it illustrates
that the scalar-matter exchange must be retained even though it affects
$r_s$, $D_M$ and $\theta_s$ only at the per-mille level, the scalar being
utterly subdominant in $\mathcal A(z_\star)$.
We note that the accumulated excursion is of the same order as the late-time
one, $|\Delta\varphi|/M\simeq0.08$, so that the framework naturally produces
field displacements of the magnitude required.

A second, independent limitation follows from the same closure condition.
Because $\eta$ is fixed by Eq.~\eqref{eq:closure} at $z=0$ rather than fitted,
the total physical matter density is an output of the construction. For the
$z_E=10$ solution one obtains $(1+\eta)\omega_b=0.1488$, and, including the
conformal factors frozen in before recombination, an effective
pre-recombination value
$\big(e^{\varphi_\star/M}+\eta\,e^{\varphi_\star/2M}\big)\omega_b=0.1509$, to
be compared with $\omega_m=0.1424$ for the reference cosmology. Since
radiation carries the same conformal factor as the baryons in
Eq.~\eqref{eq:Geff_early}, the matter--radiation equality redshift is set by
$1+z_{\rm eq}=\big(1+\eta\,e^{-\varphi_\star/2M}\big)\omega_b/\omega_r$ and is
correspondingly displaced from $z_{\rm eq}\simeq3425$ to
$z_{\rm eq}\simeq3543$. This qualifies the statement
made below Eq.~\eqref{eq:integral_constraint}, where $r_s$ was taken to follow
from densities fixed by the peak morphology: matching $\theta_s$ alone does
not enforce $\omega_m$, which the CMB constrains separately through
$z_{\rm eq}$ and the peak-height ratios. The discrepancy is monotonic in
$z_E$, with $(1+\eta)\omega_b$ equal to $0.1363$, $0.1466$, $0.1472$, $0.1488$
and $0.1629$ for the single-component ansatz and for $z_E=13$, $12$, $10$ and
$5$ respectively, so that the low-$z_E$ end of the window is disfavoured on
this ground even where it satisfies Eq.~\eqref{eq:zE_bound}. Combining the two
requirements leaves $z_E\simeq12$--$13$ as the preferred region. We do not
attempt to correct for this here, since the accompanying shift of
$G_{\rm eff}$ modifies the peak morphology as well and the two effects must be
assessed jointly in a Boltzmann code; we note it as a further reason why the
likelihood analysis anticipated in Sec.~\ref{conclusions} must treat
$\omega_m$ as a constrained rather than a derived quantity.

The compensation is not without cost. Since $\theta_s$ constrains a single
integral of $\tilde H^{-1}$, whereas baryon acoustic oscillations constrain
$\tilde H(z)$ pointwise, residual deviations remain at intermediate
redshifts. For the $z_E=10$ solution, and using the drag-epoch scale $r_d$ of
Eq.~\eqref{eq:rd} consistently in both the model and the reference
cosmology, we find that $D_H/r_d$ deviates from the $\Lambda$CDM prediction
by $-3.4\%$, $+0.7\%$, $+3.1\%$ and $+3.2\%$ at $z=0.51$, $0.93$, $1.49$ and
$2.33$ respectively, while $D_M/r_d$ deviates by $-5.2\%$, $-3.7\%$,
$-2.1\%$ and $-0.9\%$ at the same redshifts. These are larger than current DESI
uncertainties \cite{Adam1,Adam2}, and a full likelihood analysis including
BAO, supernovae and CMB distance priors is therefore required before any
claim of quantitative agreement can be made. We regard the present analysis
as identifying the structure that such a fit must have, namely a
non-monotonic scalar velocity with a modest pre-recombination component,
rather than as establishing viability.

\section{Dark-energy sector and phenomenological implications}
\label{Darkenergysector}

In the present section we show that the scalar field responsible for the
modification of the expansion rate also naturally
provides a dynamical dark-energy sector. Hence, the same scalar
degree of freedom that contributes to the alleviation of the
Hubble tension can simultaneously drive the late-time accelerated
expansion of the universe.

\subsection{Scalar potential and evolving dark energy}

The scalar-field energy density and pressure are given by
$
\rho_\varphi
=
\frac12\dot\varphi^2+V(\varphi),
$
and
$
p_\varphi
=
\frac12\dot\varphi^2-V(\varphi),
$
respectively, and thus the corresponding equation-of-state parameter is
therefore
\begin{equation}
w_\varphi
=
\frac{\frac12\dot\varphi^2-V(\varphi)}
{\frac12\dot\varphi^2+V(\varphi)}
=
\frac{\frac12M^2H_0^2f^2-V(\varphi)}
{\frac12M^2H_0^2f^2+V(\varphi)},
\end{equation}
where in the second equality we used the definition \eqref{eq:f-def}. Note
that $M^2H_0^2f^2$ carries mass dimension four and may therefore consistently
be compared with the potential energy density. For the phenomenological
ansatz introduced in the previous section, the scalar kinetic
contribution remains subdominant during most of the cosmological
evolution. Consequently, the late-time dynamics are mainly
controlled by the potential term.

The scalar evolution satisfies Eq.~\eqref{scalarfield}, with $|\varphi|\ll M$,
and therefore, once the scalar evolution is specified, the
potential can in principle be reconstructed from the desired
dark-energy behaviour. In particular, we find
\begin{equation}
V(z)
=
\rho_\varphi(z)
-
\frac12M^2H_0^2f(z)^2.
\end{equation}
Hence, an observationally motivated form of $w_\varphi(z)$ determines
the corresponding scalar potential, while conversely a given
potential specifies the resulting cosmological evolution.

Finally, we mention that for the parameter range considered in this work, the
kinetic contribution remains subdominant, and therefore the present-day
dark-energy density is effectively determined by the scalar
potential value
\begin{equation}
V(\varphi_0)\sim
(10^{-3}\,\mathrm{eV})^4,
\end{equation}
which corresponds to the characteristic energy scale associated
with late-time cosmic acceleration.

\begin{figure*}[ht!]
\centering
\includegraphics[width=0.99\textwidth]{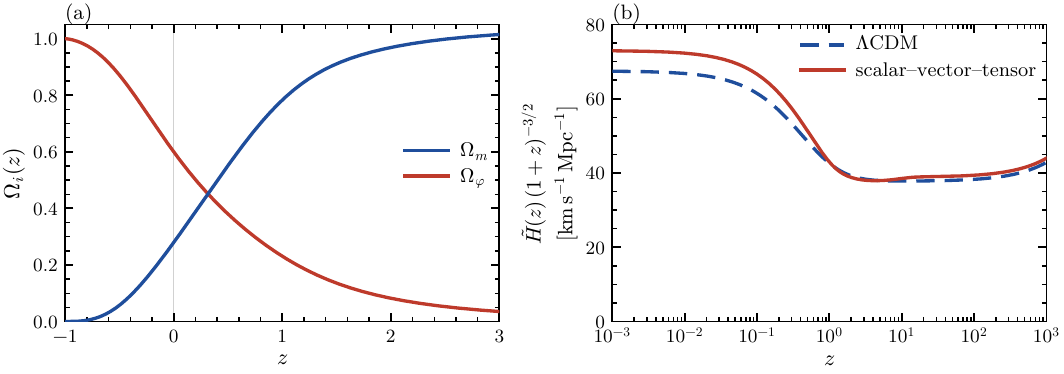}
\caption{{\it{
Left panel: Evolution of the normalized matter density parameter and scalar-field
(dark-energy) density parameter as functions of the redshift for the
scalar-vector-tensor model. The scalar-field contribution becomes
dominant only at late times, leading to the accelerated expansion of
the Universe. For the plot we have used
$\tilde{H}_0 = 73 \,\mathrm{km\,s^{-1}\,Mpc^{-1}}$ with
$\gamma=H_0/\tilde H_0$ fixed by Eq.~\eqref{eq:gamma_f0},
the two-epoch ansatz \eqref{eq:f_two_epoch} with $\alpha_L = 0.4$,
$z_L = 0.5$, $\alpha_E=1.06402$, $z_E=10$, and
$\eta = 5.65361$ (the total vector fraction, dominated by the condensate),
while for the current density parameters we have set
$\Omega_{b0} = 0.0420$ (i.e. $\omega_b=0.02237$),
$\Omega_{r0} = 7.8\times10^{-5}$, and $\bar\Omega_{\varphi0} =0.60$. The scalar
normalization scale is fixed to $M=\Mpl$, and $\varphi_0=0$. The density
parameters are those of Eq.~\eqref{eq:Ob} and obey the sum rule
\eqref{eq:sumrule}: they tend to $\Omega_\varphi\to1$, $\Omega_m\to0$ for
$z\to-1$ and to $\Omega_m\to1$ at high redshift, where $f\to0$, while in
between the sum departs from unity by the amount carried by the scalar
velocity, being $\gamma^2\simeq0.88$ today and slightly exceeding unity
around $z\simeq2$--$3$, where the early component of Eq.~\eqref{eq:f_two_epoch}
renders $f$ negative. 
Right panel: Evolution of the normalized Hubble expansion rate for the
scalar-vector-tensor model compared with the standard
$\Lambda$CDM cosmology, for the same parameters as in
left panel. For convenience, we introduce the 
dimensionless scalar variable $\phi(z)\equiv\varphi(z)/M$, normalized so
that $\phi_0=0$; the accumulated excursion over the plotted range is
$|\Delta\phi|\simeq0.08$, and the value frozen in before recombination is
$\phi_\star=+0.024$, corresponding to $r_d$ smaller than its $\Lambda$CDM
value by $2.0\%$.}}}
\label{Omegas}
\end{figure*}

In order to present our results more transparently, we evolve the cosmological
equations numerically, focusing on the behavior of the matter and the
effective dark energy density parameters $\Omega_m$ and
$\Omega_{DE}\equiv\Omega_\varphi$ defined in Eqs.~\eqref{eq:Ob}--\eqref{eq:Ophi}.
In Fig. \ref{Omegas} we depict their
evolution as a function of the redshift.
As we can see, the scenario at hand can reproduce the standard thermal history
of the Universe, with the transition from matter domination to scalar-field
domination.

Now, in order to show explicitly how the scenario at hand affects the Hubble
tension, in Fig.~\ref{Omegas} (right panel) we depict the evolution of the normalized
Hubble expansion rate. As we can see, in the scenario at hand the Hubble
function approaches that of $\Lambda$CDM at high redshift, while it is
increased at late times, due to the conformal scalar coupling. The residual
early-time offset is the $2.4\%$ shift of $G_{\rm eff}$ discussed in
Sec.~\ref{subsec:two_epoch}, which is precisely what allows the acoustic
angular scale to be preserved.

An interesting feature of the scenario is the connection between
the late-time enhancement of the Hubble expansion rate and the
properties of the dark-energy equation of state. At sufficiently
late times the scalar field evolves slowly and the Klein-Gordon equation
reduces to
\begin{equation}
3H\dot\varphi\simeq -V_{,\varphi}-\frac{\rho_b}{2M},
\label{eq:slowroll}
\end{equation}
where the last term is the conformal source already retained in
Eq.~\eqref{eq:scalar_continuity}. It is not negligible today: with
$\bar\Omega_{b0}=0.042$ and $f_0\simeq0.13$ its ratio to $3H\dot\varphi$ is
$\bar\Omega_{b0}/(2\gamma^2f_0)\simeq0.18$. Evaluating
Eq.~\eqref{eq:slowroll} at the present epoch and using
Eq.~\eqref{eq:f-def} we obtain
$
V_{,\varphi}(\varphi_0)
\simeq
-3MH_0^2f_0\left[1+\bar\Omega_{b0}/(2\gamma^2f_0)\right]
\simeq-3.5\,MH_0^2f_0,
$
which is of mass dimension three as required. The source affects the
reconstructed \emph{slope} of the potential but not $w_{\varphi0}$, which
follows from $\rho_\varphi$ and $\dot\varphi$ alone.
The present-day scalar equation-of-state parameter is
\begin{equation}
w_{\varphi0}
=
\frac{\frac12M^2H_0^2f_0^2-V_0}
{\frac12M^2H_0^2f_0^2+V_0}.
\end{equation}
Hence, in the regime where the scalar kinetic contribution remains much
smaller than the potential term, namely
$
M^2H_0^2f_0^2\ll V_0,
$
one obtains approximately
\begin{equation}
w_{\varphi0}
\simeq
-1+\frac{M^2H_0^2f_0^2}{V_0}.
\end{equation}
Therefore, the deviation from a pure cosmological constant is
controlled by the same quantity responsible for the
enhancement of the Hubble parameter. For the parameter
range relevant to the Hubble-tension mechanism,
i.e. $
f_0\sim {\cal O}(0.1)$,
and using $V_0\simeq3\Mpl^2\tilde H_0^2\bar\Omega_{\varphi0}$ together with
$M=\Mpl$, one obtains
\begin{equation}
w_{\varphi0}+1\simeq\frac{\gamma^2f_0^2}{3\bar\Omega_{\varphi0}}\approx0.009,
\end{equation}
namely the deviation of $w_{\varphi0}$ from $-1$ remains at the
percent level, and is a definite prediction correlated with $f_0$ that
future measurements of $w_0$ and $w_a$ can test.

\section{Screening mechanism and local constraints}
\label{sec:screening}

A generic feature of scalar-tensor theories is the presence of an
additional force mediated by the scalar degree of freedom. In the
present framework, matter couples to the scalar sector through the
conformal factor $e^{\varphi/2M}$, with dimensionless coupling strength
$\beta=\Mpl/2M$ given in Eq.~\eqref{eq:beta}, and therefore one expects
deviations from standard gravitational dynamics unless an appropriate
screening mechanism operates in high-density environments.

In the presence of non-relativistic matter, the scalar field experiences
an effective potential of the form
\begin{equation}
V_{\rm eff}(\varphi)
=
V(\varphi)+e^{\varphi/2M}\tilde\rho_m,
\label{eq:Veff}
\end{equation}
where $\tilde\rho_m$ is the Jordan-frame matter density.
If the potential is such that the effective potential develops a
density-dependent minimum, the scalar field acquires there an
effective mass
\begin{equation}
m_\varphi^2
=
\frac{d^2V_{\rm eff}}{d\varphi^2},
\end{equation}
which is of mass dimension two, since $[V]=4$ and $[\varphi]=1$.
Consequently, in high-density environments such as the Solar
System, the scalar interaction becomes short-ranged and the
corresponding fifth force is suppressed, while at
cosmological densities the scalar field remains light and can
affect the expansion history.

Two caveats must be stated explicitly. First, the matter contribution to
Eq.~\eqref{eq:Veff} is monotonic in $\varphi$, so that a minimum exists only
if $V(\varphi)$ is of runaway type; a slowly rolling quintessence-like
potential of the kind reconstructed in Sec.~\ref{Darkenergysector} does not
by itself produce one. A chameleon therefore requires supplementing
$V$ with a runaway piece, for instance $V\supset \Lambda^{4+n}\varphi^{-n}$,
in which case $V_{,\varphi\varphi}$ rather than the matter term dominates
$m_\varphi^2$ at the minimum. Second, retaining only the matter term,
\begin{equation}
m_\varphi^2\sim \frac{e^{\varphi/2M}\tilde\rho_m}{4M^2}
=
\frac{\beta^2 e^{\varphi/2M}\tilde\rho_m}{\Mpl^2},
\label{eq:mphi_matter}
\end{equation}
one finds for $M=\Mpl$ and terrestrial densities an interaction range
$m_\varphi^{-1}\sim10^{11}\,$m, four orders of magnitude larger than the
Earth radius, so that no thin shell forms and the fifth force is
\emph{not} screened. We therefore emphasize that the choice $M=\Mpl$ adopted
in the numerical analysis is viable only if the potential supplies the
required runaway behaviour; in the absence of such a term, local tests demand
$M\gtrsim2\times10^{2}\Mpl$ as obtained from Eq.~\eqref{eq:gammaPPN}. Since
the cosmological analysis of Secs.~\ref{sec:cosmo}--\ref{latetime} depends on
$M$ only through the combination $\varphi/M$, this rescaling can be absorbed
without affecting the background results, at the price of a correspondingly
larger $\varphi$.

At weak-field scales, the scalar contribution to the gravitational
potential can be parametrized as
\begin{equation}
\Phi_\varphi(r)
=
-\alpha_\varphi
\frac{GM_b}{r}
e^{-m_\varphi r},
\end{equation}
where $\alpha_\varphi$ characterizes the strength of the scalar
interaction relative to Newtonian gravity. We emphasize that
$\alpha_\varphi$ is not a free parameter but is fixed by the normalization
scale $M$ through
\begin{equation}
\alpha_\varphi=2\beta^2=\frac{\Mpl^2}{2M^2},
\label{eq:alphaphi}
\end{equation}
so that $\alpha_\varphi=1/2$ for the choice $M=\Mpl$. An
unscreened scalar of this strength would be excluded by Solar-System tests,
and the screening mechanism described above is therefore an essential
ingredient of the construction rather than an optional refinement.
Additionally, the vector sector contributes an additional Yukawa-type
correction of the form
\begin{equation}
\Phi_v(r)
=
+\alpha_v
\frac{GM_b}{r}
e^{-m_v r},
\,\,
\alpha_v=\frac{g_v^2}{4\pi G m_b^2}=\frac{2g_v^2\Mpl^2}{m_b^2},
\label{eq:alphav}
\end{equation}
where the positive sign corresponds to a repulsive contribution and
$\alpha_v$ is dimensionless, as follows from $[g_v]=0$.
Combining the standard Newtonian term with the scalar and vector
contributions, the total weak-field potential becomes
\begin{equation}
\Phi_{\rm tot}(r)
=
-\frac{GM_b}{r}
\left[
1+\alpha_\varphi e^{-m_\varphi r}
-\alpha_v e^{-m_v r}
\right].
\label{eq:totalpotential}
\end{equation}

In sufficiently dense environments the scalar interaction becomes
effectively screened, $e^{-m_\varphi r}\to0$, and if in addition the vector
is short-ranged one recovers
\begin{equation}
\Phi_{\rm tot}(r)
\simeq
-\frac{GM_b}{r}.
\end{equation}
We note, however, that the vector Yukawa suppression is efficient only for
the density-dependent-mass case $s\neq0$; for the constant-mass condensate
of Sec.~\ref{subsec:CMB_vector}, the requirement $m_\star\gg H(z_\star)$ is
compatible with a Compton wavelength far exceeding Solar-System scales, in
which case local constraints must instead be satisfied by taking $\alpha_v$
itself sufficiently small. Given the estimate of $\eta_{\rm int}/\alpha_v$
obtained in Sec.~\ref{subsec:CMB_vector}, this is not restrictive for the
background cosmology, which is controlled by $\eta_{\rm c}$.

At larger galactic distances the scalar
screening can become less efficient. In particular, if the scalar
interaction range is larger than the vector one, namely for
$
m_\varphi<m_v,
$
the vector contribution becomes exponentially suppressed more
rapidly, while the scalar interaction can still provide an
additional attractive component, so that
\begin{equation}
\Phi_{\rm tot}(r)
\simeq
-\frac{GM_b}{r}
\left[
1+\alpha_\varphi e^{-m_\varphi r}
\right].
\end{equation}
Therefore, the scalar sector can effectively enhance the
gravitational attraction at galactic scales, while remaining
screened in high-density environments. In this sense, the model
contains the necessary ingredients for reproducing dark-matter-like
phenomenology at astrophysical scales, while preserving consistency
with local gravity tests.

A complete analysis of the allowed parameter space would require a
detailed treatment of the scalar and vector field profiles in
realistic astrophysical environments, together with a full
comparison against laboratory, Solar-System, and galactic
constraints, including in particular a quantitative thin-shell
analysis for the adopted potential. Such an investigation lies beyond the
scope of the present work and is left for future study.

\section{Conclusions}
\label{conclusions}

In this work we investigated a scalar-vector-tensor cosmological
framework motivated by the possibility that the expansion
history may differ mildly from the one inferred within the standard
$\Lambda$CDM scenario. The central idea was to explore whether a
conformally coupled scalar degree of freedom can induce a controlled
modification of the physical Hubble rate, while leaving recombination and
nucleosynthesis unaffected. Additionally, we included
a massive vector sector coupled to the baryonic current, which can
provide an effective pressureless contribution at the background level.

We first formulated the theory in the Einstein frame and introduced the
Jordan-frame metric to which matter is minimally coupled. The scalar sector
is normalized by a mass scale $M$, so that the canonically normalized field
$\varphi=M\ln(1+\Xi)$ carries mass dimension one and the matter coupling is
controlled by the dimensionless combination $\beta$. This allowed us
to derive the relation between the Einstein-frame Hubble parameter and
the expansion rate measured by matter observers; in the small-field regime this becomes
$
\tilde H\simeq H+\frac12H_0 f(z),
$
where $f(z)=\dot\varphi/(MH_0)$ is dimensionless.

A central point of our analysis is that, although $H_0$ is not a conformal
invariant, the acoustic angular scale $\theta_s=r_s/D_M$ is, since the
comoving separation $\int d\tilde t/\tilde a=\int dt/a$ is frame-independent.
The tension is therefore a statement between invariants and cannot be
dissolved by a change of frame. This yields the exact integral condition
$\int_0^{z_\star}dz/\tilde H=r_s/\theta_s$, which any admissible $f(z)$ must
satisfy and which a strictly positive $f(z)$ necessarily violates: for the
single-component ansatz we find $\theta_s$ shifted by $2.5\%$. We therefore
introduced a two-epoch ansatz in which $f$ changes sign, leaving a small
positive $\varphi_\star$ frozen in before recombination. This rescales the
effective gravitational coupling by $G_{\rm eff}/G=e^{\varphi_\star/M}$,
shrinking $r_s$, and we showed that $\theta_s$ can then be matched exactly
with $\varphi_\star/M\simeq0.024$, that is a $2.4\%$ shift of $G_{\rm eff}$,
while retaining $\tilde H_0=73\,\mathrm{km\,s^{-1}\,Mpc^{-1}}$. Because the
scalar exchanges energy with matter, its continuity equation carries a source
term which dominates the early epoch and drives $\rho_\varphi$ towards the
past; requiring $\bar\Omega_\varphi(z_\star)>0$ then bounds
$z_E\lesssim13$, a restriction invisible at the level of the background
distances alone. The recombination temperature is untouched, since particle
masses are constant in the Jordan frame, although the ionization history
responds weakly to the modified expansion rate. Two costs remain: residual
deviations of a few percent in the BAO observables, and a displacement of the
total matter density, $(1+\eta)\omega_b=0.1488$ against $0.1424$ for the
reference cosmology, which the CMB constrains independently of $\theta_s$. We
therefore do not claim quantitative agreement at this stage.

We further showed that the same scalar sector naturally gives rise to a
dynamical dark-energy component through its potential $V(\varphi)$. For a
slowly evolving scalar field the kinetic contribution remains
subdominant, and the dark-energy density is mainly controlled by the
potential term. Moreover, the deviation of the present dark-energy
equation-of-state parameter from $-1$, namely
$w_{\varphi0}+1\simeq M^2H_0^2f_0^2/V_0$, is directly related to the scalar
quantity that controls the Hubble-rate enhancement. Thus, the model
links the modification of the expansion rate with the
dynamics of dark energy, offering a phenomenological consistency
relation that can be tested by future measurements of $w_0$ and $w_a$.

The vector sector was shown to provide two distinct contributions. The
temporal component, determined algebraically by the baryon current,
yields an apparent matter-like term in the background expansion that is
not a true fluid but rather a manifestation of the interaction energy.
This component is locked to the baryon density, $\delta\rho_v^{(\rm int)}/\rho_v^{(\rm int)}=\delta_b$,
and therefore does not cluster independently. The propagating spatial modes,
on the other hand, form a vector condensate that behaves as a
collisionless pressureless component and can play the cosmological role of
cold dark matter. In analogy with other massive-vector cosmologies,
such modes could in principle provide an independently clustering
component if they approach $w_v\simeq c_{s,v}^2\simeq\sigma_v\simeq0$.
Whether this regime is dynamically realized in the present
density-dependent-mass theory requires a full linear perturbation and
Boltzmann analysis.

At the same time, local deviations from standard gravity must be suppressed
through a chameleon-like screening mechanism for the scalar field. We showed
that for $M=\Mpl$ this requires the scalar potential to contain a runaway
piece, since the matter term alone yields an interaction range far exceeding
the size of the source; in the absence of such a term local tests require
$M\gtrsim2\times10^2\Mpl$, which the background analysis can accommodate.

Several extensions of the present analysis are required in order to
assess the full viability of the scenario. A complete statistical
confrontation with cosmological data, including CMB distance priors,
BAO measurements, supernovae, cosmic chronometers, and local $H_0$
priors, will determine the allowed parameter space and quantify the
degree of Hubble-tension alleviation, treating $M$ (equivalently $\beta$),
$\alpha_L$, $\alpha_E$, $z_L$ and $z_E$ as free parameters subject to the
constraint \eqref{eq:integral_constraint}, to the positivity requirement
\eqref{eq:zE_bound}, and to a prior on $\omega_m$ rather than to $\theta_s$
alone. Furthermore, a detailed study of
linear perturbations and growth observables, such as $f\sigma_8$ and
$S_8$, is necessary in order to test the model beyond the background
level. Finally, a more complete investigation of the screening
mechanism and of the scalar/vector profiles in realistic astrophysical
environments will be essential for confronting the theory with
Solar-System, laboratory, and galactic-scale constraints. These studies are
left for future works.

\acknowledgments{A.A.K and A.S. thank Shiraz university Research
Council. The authors acknowledges the contribution of COST Actions CA21106 
``COSMIC WISPers in the Dark Universe: Theory, astrophysics and experiments'',  
CA21136 ``Addressing observational tensions in cosmology with systematics and 
fundamental physics (CosmoVerse)'', CA23130 ``Bridging high and low energies 
in search of quantum gravity (BridgeQG)'', and CA24101 ``Testing Fundamental 
Physics with Seismology''. JLS would also like to acknowledge funding from 
``Xjenza Malta'' as part of the ``Technology Development Programme'' 
DTP-2024-014 (CosmicLearning) Project.
}

\end{document}